\documentclass[12pt]{article}

\usepackage[margin = 2cm]{geometry}
\usepackage{graphicx}
\usepackage{amssymb}
\usepackage{lineno}
\usepackage[title]{appendix}

\usepackage{multirow}
\usepackage{setspace}
\usepackage{natbib}
\usepackage{booktabs}
\usepackage{caption}
\usepackage{amsmath}
\usepackage{hyperref}
\usepackage{ccicons}
\usepackage[T1]{fontenc}
\usepackage{enumitem}
\usepackage[table,xcdraw]{xcolor}
\usepackage{floatrow}
\usepackage{subcaption}
\usepackage{authblk}
\usepackage{booktabs}
\usepackage[normalem]{ulem}
\useunder{\uline}{\ul}{}
\graphicspath{ {././} }

\begin{document}

\title{Attention-based dynamic multilayer graph neural networks for loan default prediction \footnote{\scriptsize NOTICE: This is the author’s version of a submitted work for publication. Changes resulting from the publishing process, such as editing, corrections, structural formatting, and other quality control mechanisms may not be reflected in this document. Changes may have been made to this work since it was submitted for publication. This work is made available under a \href{https://creativecommons.org/licenses/by/4.0/}{Creative Commons BY-NC-ND license}. \ccbyncnd}}

\author[1]{Sahab Zandi}
\author[2,3]{Kamesh Korangi}
\author[4]{Mar\'{i}a \'{O}skarsd\'{o}ttir}
\author[2,3]{Christophe Mues}
\author[1]{Cristi\'{a}n Bravo}

\affil[1]{Department of Statistical and Actuarial Sciences, Western University, 1151 Richmond Street, London, Ontario, N6A 5B7, Canada}
\affil[2]{Department of Decision Analytics and Risk, Southampton Business School, University of Southampton, University Road, SO17 1BJ, United Kingdom}
\affil[3]{Centre for Operational Research, Management Sciences and Information Systems, University of Southampton, University Road, SO17 1BJ, United Kingdom}
\affil[4]{Department of Computer Science, Reykjav\'{i}k University, Menntavegur 1, 102 Reykjav\'{i}k, Iceland}

\date{}
\maketitle

\begin{abstract}
Whereas traditional credit scoring tends to employ only individual borrower- or loan-level predictors, it has been acknowledged for some time that connections between borrowers may result in default risk propagating over a network. In this paper, we present a model for credit risk assessment leveraging a dynamic multilayer network built from a Graph Neural Network and a Recurrent Neural Network, each layer reflecting a different source of network connection. We test our methodology in a behavioural credit scoring context using a dataset provided by U.S. mortgage financier Freddie Mac, in which different types of connections arise from the geographical location of the borrower and their choice of mortgage provider. The proposed model considers both types of connections and the evolution of these connections over time. We enhance the model by using a custom attention mechanism that weights the different time snapshots according to their importance. After testing multiple configurations, a model with GAT, LSTM, and the attention mechanism provides the best results. Empirical results demonstrate that, when it comes to predicting probability of default for the borrowers, our proposed model brings both better results and novel insights for the analysis of the importance of connections and timestamps, compared to traditional methods.
\end{abstract}


\section{Introduction}
\label{Introduction}

Network science provides a beneficial tool to study complex systems of interacting entities that can be found in many areas, such as biology, finance and economics \citep{barabasi2016network}. To represent connections between these entities, graphs are a common representation method, which have diverse applications in social network analysis \citep{haythornthwaite1996social}, computational finance \citep{wang2021review}, and recommender systems \citep{wang2021graph}, among many others. Graph Neural Networks (GNNs) are models in the field of deep learning, specifically tailored to perform over graph domains. They have been utilized for different tasks, including node classification \citep{tang2021graph}, edge prediction \citep{zhang2018link}, and graph clustering \citep{tsitsulin2020graph}. In most cases, they have been used with static single layer networks, in which nodes are linked based on one source of connection, and the network remains unchanged over time. In reality, though, nodes could be connected by more than one source of connection, as is the case in our application setting. Such networks are generally called multilayer networks \citep{kivela2014multilayer}. Furthermore, the nodes and edges in a graph may evolve. For example, new nodes may appear, node features may change, and a new relation may emerge between two nodes. Being able to capture these changes in the models could lead to higher predictive performance for problems that are characterized by such dynamic graphs.

To enable dynamic graph learning, we consider Recurrent Neural Networks (RNNs) \citep{elman1990finding}. RNNs are used for data that is presented in a sequence, such as time series data or natural language. Their main objective is to create a representation of a series of inputs, usually indexed by time, to predict an output. Because of their powerful learning capacity, they have been applied successfully in various types of tasks, including speech recognition \citep{graves2007multi}, acoustic modelling \citep{qu2017syllable}, trajectory prediction \citep{altche2017lstm}, sentence embedding \citep{palangi2016deep}, and correlation analysis \citep{mallinar2018deep}. Since dynamic graphs represented by discrete snapshots can be considered sequence data, RNNs provide a solution to capture the evolution of these graphs. However, it is known that attention-based RNNs outperform encoder-decoder-based RNNs, indicating that the incorporation of attention can improve the prediction performance \citep{aliabadi2020attention}.

The application area that we focus on in this paper is credit scoring --- one of the prominent applications of data analytics. Lenders build credit scoring models to help adjudge the risk involved in granting a loan and decide on the terms of the loan and the interest rate \citep{thomas2017credit}. Traditional credit scoring models use loan- or borrower-level data to assess a loan applicant's default risk, thus treating borrowers as independent entities. While potential default correlation between borrowers has been acknowledged for some time, it is only more recently that this has started to be further investigated using network science. This is part of a broader trend in which credit scoring research increasingly focuses on improving the performance of existing credit scoring models through the incorporation of machine learning methods, and the inclusion of alternative data sources such as network data \citep{bravo2020evolution}.

In this paper, we propose using an attention-based dynamic multilayer graph neural network to model the problem of credit risk across time and explicitly incorporate default correlation between borrowers. This work makes the following contributions. Firstly, our solution, Dynamic Multilayer Graph Neural Networks (DYMGNN), represents a novel approach for node classification in multilayer networks. Secondly, we show how to apply the proposed method to credit risk modelling, using the example context of mortgage loan default prediction. Finally, we show that, in this setting, our model, by considering dynamicity, multilayer effects, and using an attention mechanism, outperforms other baseline methods.

The structure of this paper is as follows. The next section discusses a selection of previous work on GNN and credit risk modelling. Section~\ref{Methodology} explains the methodology, multilayer networks, embeddings, and the models used in this paper. Section~\ref{Experimental setup} sheds light on the data, dynamic networks, and the experiments in the paper. Section~\ref{Results and discussion} presents the experimental results and highlights some discussion points relevant to the models. The final section summarizes our conclusions and suggests future work.

\section{Previous work}
\label{Previous work}

\subsection{Graph neural networks}
\label{Graph neural networks}
Graphs can be seen in many real-world applications. In some cases, the graph is static, i.e., the graph structure and node features do not change over time. In other cases, the graph is dynamic, i.e., the graph evolves. GNNs are neural models that capture the dependencies between the nodes within a graph through message passing between the nodes of the graph. A comprehensive survey of methods and applications related to GNNs is provided by \citet{zhou2020graph}. Recently, some types of GNNs such as the Graph Convolutional Network (GCN) \citep{kipf2016semi} and Graph Attention Network (GAT) \citep{velivckovic2018graph} have been widely used for various deep learning tasks, albeit mostly on static graphs.

GCN is an approach for semi-supervised learning on graph-structured data, utilizing an efficient layer-wise linear model based on a first-order approximation of spectral graph convolutions. To prevent overfitting, it simplifies the convolution calculation by constraining the number of parameters, and by minimizing the number of operations, for example, reducing the matrix multiplications per layer. The number of graph edges is linearly scaled in GCN and this model learns hidden layer representations in which both the local graph structure and node features are encoded. GCN has shown acceptable performance on citation networks and knowledge graphs \citep{kipf2016semi}.

A second type of GNN, known as GAT, uses an attention mechanism to learn node-level representations \citep{velivckovic2018graph}. The encoder-decoder-based neural machine translation system was outperformed by the attention mechanism in natural language processing \citep{bahdanau2014neural}. Nowadays, attention models are widely utilized for document categorization \citep{pappas2017multilingual}, recommendation systems \citep{xiao2017attentional}, and the creation of image captions \citep{xu2015show}. The attention mechanism concentrates on a few selected relevant attributes while ignoring other irrelevant attributes \citep{bahdanau2014neural}. The categories of attention models that are now in use are global and local attention \citep{luong2015effective}, soft and hard attention \citep{bahdanau2014neural}, and self-attention \citep{vaswani2017attention}. Global attention considers all source positions when deriving the context vector, providing a comprehensive overview of the entire input sequence. In contrast, local attention focuses on a subset of source positions, making it more efficient by narrowing the context to relevant parts. Soft attention generates a weighted sum of the attention scores, allowing gradients to pass through and making the model end-to-end trainable. Hard attention, on the other hand, selects a single source position, making it less computationally expensive but more challenging to optimize due to its non-differentiable nature. Self-attention is an attention mechanism that is applied to compute a representation of a single sequence, enabling the model to weigh the importance of different positions in the sequence when generating its output. In conjunction with RNN or convolutions, self-attention can be used in many applications like learning sentence representations \citep{lin2017structured} and machine reading \citep{cheng2016long}. The GAT is a type of GNN that applies an attention mechanism to graph-structured data, so as to classify nodes. It computes the hidden representation of each node by paying attention to its adjacent nodes and then applying a self-attention strategy. GAT achieved state-of-the-art results in both transductive (semi-supervised) and inductive (supervised) settings.

Most of the GNN models have been proposed for static graph learning; however, over the past few years, several machine learning models capturing the structure and evolution of dynamic graphs have been introduced. A complete review of representation learning approaches for dynamic graphs is given in \citet{kazemi2020representation}, while a more specialized review of GNN-based approaches for dynamic graphs is provided by \citet{skarding2021foundations}. There are many different approaches to modelling spatial and temporal information in graph-structured data. Diffusion Convolutional Recurrent Neural Network \citep[DCRNN;][]{li2017diffusion} and Spatio-Temporal Graph Convolutional Networks \citep[STGCN;][]{yu2017spatio} analyse graph-structured data first and pass the results to sequence-to-sequence models or RNNs. Structural-RNN collects spatial and temporal data synchronously to associate the graph structure with temporal data so that it can apply RNNs on new graphs \citep{jain2016structural}. Dynamic Graph Convolutional Networks (DGCN) propose a novel approach that combines RNNs and GCNs to learn long short-term dependencies together with the graph structure \citep{manessi2020dynamic}. EvolveGCN is an approach for graph representation learning in dynamic graphs that uses an RNN to evolve the parameters of a GCN model. By doing so, EvolveGCN can capture the dynamics of the graph sequence without relying on node embeddings \citep{pareja2020evolvegcn}. Temporal Graph Convolutional Network (T-GCN) is a novel method for real-time traffic forecasting that uses GCN to learn the complex topological structure of the urban road network for spatial dependence. It also employs RNN to capture the dynamic changes in traffic data for temporal dependence \citep{zhao2020t}. Temporal Graph Attention (TGAT) has been proposed for inductive representation learning on temporal graphs that uses a self-attention mechanism and a novel functional time encoding technique to efficiently aggregate temporal-topological neighbourhood features and learn time-feature interactions. TGAT can handle both node classification and link prediction tasks, and can be extended to include temporal edge features \citep{xu2020inductive}. Joint Dynamic User-Item Embeddings (JODIE) has been proposed to predict future user-item interactions in domains such as e-commerce, social networking, and education. It is a coupled recurrent neural network model that learns the embedding trajectories of users and items through representation learning, and it introduces a novel projection operator to estimate the embedding of the user at any time in the future \citep{kumar2019predicting}. Dynamic Representation (DyRep) encodes evolving information over dynamic graphs into low-dimensional representations, namely as embeddings, using an inductive deep representation learning framework. The learned embeddings drive the dynamics of two fundamental processes: communication and association between nodes in dynamic graphs \citep{trivedi2018dyrep}. Dynamic Self-Attention Network (DySAT) computes node representations by jointly using self-attention layers along two dimensions: structural neighbourhood and temporal dynamics, and has been evaluated on link prediction experiments \citep{sankar2018dynamic}.

The methods mentioned above are all designed for single layer networks, and would thus not be suitable for the multilayer problem we tackle. Recently, however, several approaches to generalize GNNs to the multilayer case have been proposed. Firstly, Graph Attention Models for Multilayered Embeddings (GrAMME) introduces attention mechanisms and develops two GNN architectures to exploit the interlayer dependencies: GrAMME-SG and GrAMME-Fusion. GrAMME-SG considers a multilayer network to be a supra graph with implicit edges between layers, whereas GrAMME-Fusion makes use of a supra fusion layer to aggregate embeddings from layerwise attention models \citep{shanthamallu2019gramme}. Secondly, Multilayer network Embedding via Learning Layer vectors (MELL) incorporates the idea of a layer vector that characterizes the connectivity in a layer. MELL embeds nodes in each layer into the lower embedding space using all layer structures and incorporates layer vectors to differentiate edge probabilities in the layers \citep{matsuno2018mell}. Thirdly, Multilayer Graph Neural Network (mGNN) presents an innovative way of employing GCN on multilayer networks. In this approach, node feature propagation occurs independently in both intralayer and interlayer edges, and multiple layers can be stacked to capture information from the topology and features further in the network. This method can handle node classification, intra layer link prediction, and network clustering \citep{grassia2021mgnn}.

All the aforementioned methods have been developed for networks that are either static and single layer, static and multilayer, or dynamic and single layer. A unified approach that can handle networks that are both dynamic and multilayer has not been put forward yet. This is significant, as most real-life networks share both characteristics simultaneously. Our approach focuses on solving this problem.

\subsection{Credit risk modelling with network data}
\label{Credit risk modelling with network data}
Credit risk modelling has a long history, and researchers from a broad range of areas have been working on developing credit risk rating systems \citep{markov2022credit}. The statistical models that were traditionally used for credit risk modelling seemed to have difficulties dealing with large datasets as they may be characterized by increased noise, heavy-tailed distributions, nonlinear patterns, and temporal dependencies \citep{gordy2000comparative}. Advances in computing power and availability of large credit datasets paved the way to artificial intelligence (AI) driven credit risk estimation algorithms such as machine learning and deep learning \citep{shi2022machine}. Recently, some machine learning methods were found to outperform conventional models in terms of accuracy when applied to large datasets \citep{lessmann2015benchmarking, gunnarsson2021deep}.

Whereas traditional methods such as logistic regression make the IID assumption, treating borrowers as independent observations, it is widely understood that correlated default exists in lending \citep{oskarsdottir2021multilayer}. Default correlation measures the extent to which the default of one borrower is related to that of another borrower, which may be caused by similar economic conditions affecting both, or, within a sector, by industry-specific reasons. Several researchers have shown that these correlations should be taken into account to avoid misestimating credit risk \citep{fenech2015loan}. Some researchers have used alternative data sources such as network data to show the existence of correlation. For this purpose, they used different sources of data such as telephone call data \citep{oskarsdottir2019value}, app-based marketplace data \citep{roa2021super}, social media data \citep{de2019does}, and agricultural loan network data \citep{oskarsdottir2021multilayer}. Some other researchers have used network data from inter-firm transactions to improve credit risk modelling strategies for Small and Medium-sized Enterprises (SMEs) \citep{vinciotti2019effect}. A common feature of the aforementioned studies is that they all use numerical measures obtained from the alternative data sources. However, all of them consider networks that are either static or single layer. In fact, nodes could be connected in various ways in a network, and these relationships could change over time, making them dynamic. Our work is an effort to take advantage of dynamic multilayer networks in this field and thus address some of the limitations of previous work.

Whereas traditionally numerical features were extracted from the network and used for credit scoring, more recently, GNNs have been used to develop predictive models in this field, eliminating the need to explicitly extract those features. In one study, GCNs were employed to predict peer-to-peer loan defaults and were found to outperform baseline models such as SVM, Random Forest, and XGBoost \citep{lee2021graph}. GNN with Self-attention and Multi-task learning (SaM-GNN) has been proposed for credit default risk prediction. This approach incorporates two parallel tasks based on shared intermediate vectors for input vector reconstruction and credit default risk prediction \citep{li2022graph}. Motif-preserving Graph Neural Network with curriculum learning (MotifGNN) has been introduced to jointly learn the lower-order structures from the original graph and higher-order structures from multi-view motif-based graphs for default prediction \citep{wang2023financial}. In another study, a novel spatial-temporal aware GNN was proposed to predict SME loan default risk from a network of mined supply chain relationships \citep{yang2021financial}. The potential benefits of using networks that are both dynamic and multilayer in credit risk modelling are not yet fully realized. This research presents a novel method for utilizing these networks within this area of study.

\section{Methodology}
\label{Methodology}

In this section, we describe our methodology. First, we explain our approach for constructing a sequence of snapshots of multilayer networks. Then, we discuss how we employ different types of embeddings to encode topological and temporal dependencies in the networks. After that, we describe different configurations of DYMGNN and their respective architectures.

\subsection{Multilayer networks}
\label{Multilayer networks}
\begin{figure}[hbt!]
\includegraphics[scale=0.9]{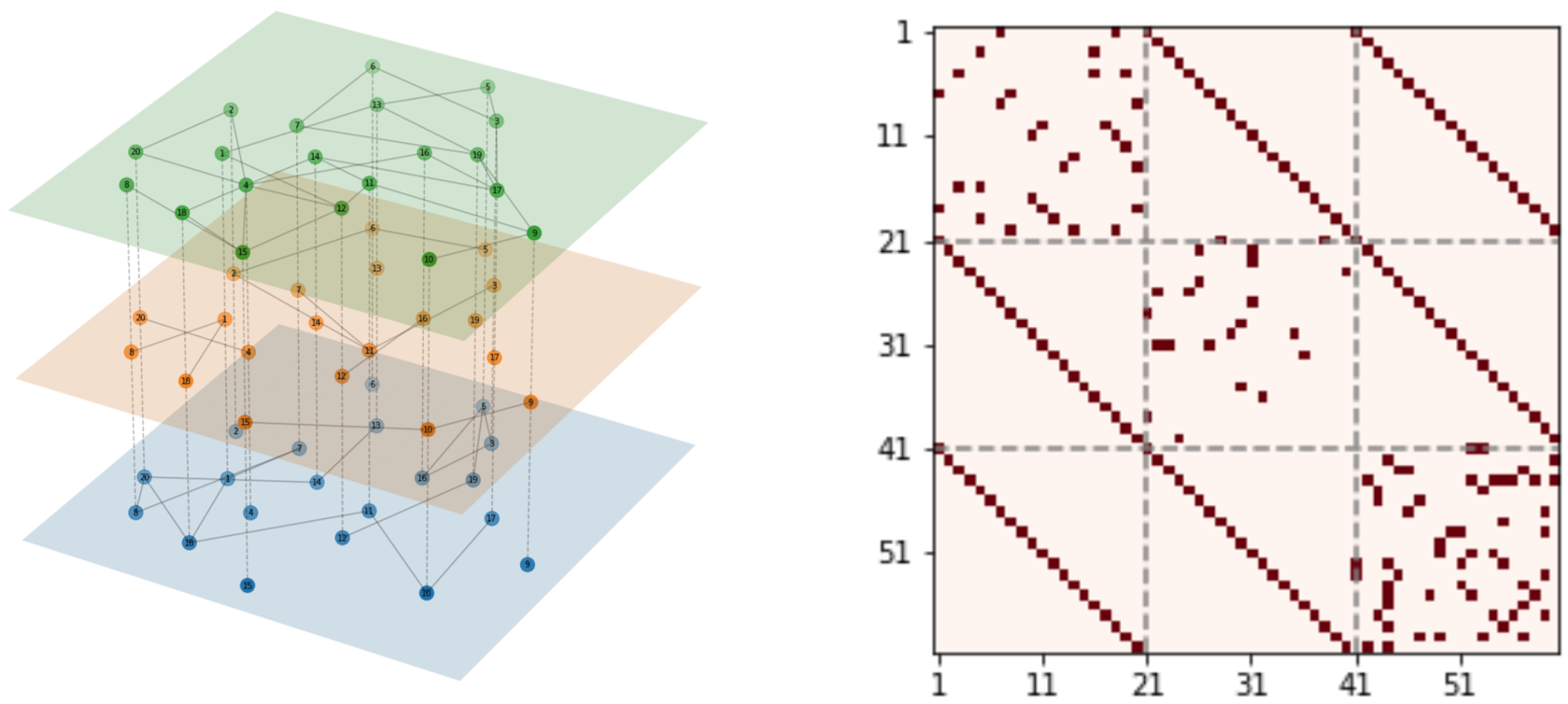}
\centering
\caption{A multilayer network (left) and its supra adjacency matrix (right).}
\label{fig1}
\end{figure}

We start by defining the multilayer network and its node features. Consider an unweighted and undirected network $G=(V,A,X)$, where $V=\{v_1, v_2,...,v_n\}$ is the set of nodes in a layer of the network, $n=|V|$ denotes the number of distinct nodes, and $X\in\mathbb{R}^{n \times d}$ is a feature matrix, in which $X_i$ is a column vector that represents the features of node $v_i$ and $d$ stands for the number of features. A network is represented by its supra adjacency matrix, $A\in\mathbb{R}^{nl \times nl}$, where $l$ denotes the number of layers. This matrix encodes information about connections between pairs of nodes within a layer as well as connections between pairs of nodes from two different layers, i.e., if $v_i$ from layer $k$ and $v_j$ from layer $m$ are connected $(1\leq k,m\leq l)$, then $A_{(k-1)n+i,(m-1)n+j}=1$; otherwise, the value is $0$. In a multilayer network, all layers contain the same set of nodes, while their edge sets are assumed to be different. Each layer focuses on a particular type of relationship, with intra layer edges connecting the nodes that are related. In addition, a series of interlayer edges simply specify which nodes are identical. Fig.~\ref{fig1} shows a multilayer network and its corresponding supra adjacency matrix.

The network dynamics are captured through a sequence of snapshots $[G^{(1)},...,G^{(\tau)}]$ where $G^{(t)}=(V^{(t)},A^{(t)},X^{(t)})$ for each $t\in \{1,...,\tau\}$. We are interested in obtaining the node embeddings at $t\leq \tau$ based on snapshots at or before $t$. For the application in this paper, we assume $V^{(1)}=V^{(2)}=...=V^{(\tau)}$ and $A^{(1)}=A^{(2)}=...=A^{(\tau)}$, i.e., nodes and their connections remain constant over time, but features can vary from one snapshot to another. Once the set of multilayer networks is complete, an embedding must be calculated from this data, as explained in the next section.

\subsection{Topological embedding}
\label{Topological embedding}
Neural networks are a type of machine learning model inspired by the human brain, consisting of layers of interconnected neurons. Each neuron processes input data, applies a mathematical transformation using weights and biases, and passes the result to the next layer. They are trained through a process called backpropagation, which adjusts the weights and biases to minimize prediction errors measured by a loss function \citep{GoodBengCour16}. GNNs extend these neural networks to handle graph-structured data.

Capturing the topological dependence in a network is a key problem, as neighbouring nodes could influence each other. In this work, we trial two different types of GNNs, i.e., GCN \citep{kipf2016semi} and GAT \citep{velivckovic2018graph}, to obtain the topological relationship between a node and its neighbours, encode the topological structure of the network and the features of nodes, capturing the information within the node connections. GCN or GAT is applied to each $G^{(t)}$ to obtain a hidden representation matrix $Z^{(t)}$. Each row of $Z^{(t)}$ contains a node embedding, meaning that for node $v_i$ we have a sequence of embeddings $[Z_i^{(1)},Z_i^{(2)},...,Z_i^{(\tau)}]$.

The GCN formulation performs isotropic aggregation, according to which each neighbour contributes equally to update the representation of the central node. The GCN model for a snapshot can be expressed as follows:
\begin{eqnarray}
&Z=\Tilde{D}^{-1/2}\Tilde{A}\Tilde{D}^{-1/2}XW^T.\label{eq1}
\end{eqnarray}

Here, $\Tilde{A}=A+I_{nl}$ is the supra adjacency matrix of the snapshot with inserted self-loops. $I_{nl}$ is the identity matrix, $\Tilde{D}_{ii}=\sum_{j}\Tilde{A}_{ij}$ is the diagonal degree matrix of $\Tilde{A}$, and $W^T \in \mathbb{R}^{d \times D}$ is a learnable weight matrix where $D$ is the embedding size.

The GAT model expands the basic aggregation function of the GCN, assigning different importance to each edge through the attention coefficients. It can be formulated as follows:
\begin{eqnarray}
&e_{ij}=\textrm{LeakyReLU}(a^T[WX_i||WX_j]),\label{eq2}\\[10pt]
&\alpha_{ij}=\frac{exp(e_{ij})}{\sum_{k\in {N(v_i)}\cup\{v_i\}}exp(e_{ik})},\label{eq3}\\[10pt]
&Z_i=\sum_{j\in N(v_i)\cup\{v_i\}}\alpha_{ij}WX_j.\label{eq4}
\end{eqnarray}

Equation \ref{eq2} computes a pairwise denormalized attention score between two neighbours, where $||$ denotes the concatenation operation and $a^T \in \mathbb{R}^{1 \times 2D}$ is a learnable weight vector. The attention score indicates the importance of a neighbour node in the message passing framework. Equation \eqref{eq3} applies a $softmax$ function to normalize the attention scores on each node’s incoming edges. This function puts the output of the previous step in a probability distribution and, as a result, the attention scores are more comparable across different nodes. In this equation, $N(v_i)$ represents the neighbourhood of node $v_i$. Note, we also include the self-edge for each node. In Equation \eqref{eq4}, the embeddings from neighbours are aggregated together, scaled by the attention scores. The main objective of this process is to learn a different contribution from each neighbour. The operations from \eqref{eq2} to \eqref{eq4} constitute a single head. The modelling capacity can be improved by considering multiple attention heads, thus allowing for different attention being given to different sets of neighbours. The output representations from the different heads can be aggregated using averaging operations.

\subsection{Temporal embedding}
\label{Temporal embedding}
Dealing with the temporal dependence is another key problem, as the temporal sequence of connections between nodes could provide useful information. In this work, we use LSTM \citep{hochreiter1997long} and GRU \citep{cho2014properties} to capture the information related to the evolution of the networks. Both LSTM and GRU use gated mechanisms to memorize as much information as possible; however, there are some differences between these two models \citep{chung2014empirical}. Comparing these two, LSTM has a more complex structure, more parameters, and longer training time. LSTM is known to be able to deal with long-range dependencies, making it the preferred choice for models built over data of relatively large size \citep{yang2020lstm}. As our data is of medium size and the interface between GNNs and RNNs is not fully explored, the choice between LSTM and GRU is not obvious. We will, therefore, compare both models in Section~\ref{Results and discussion}. After obtaining the sequence of topological embeddings $[Z_i^{(1)},Z_i^{(2)},...,Z_i^{(t)}]$, we need to input it into the RNN model and use the hidden representation of the RNN model as the temporal node embeddings for $v_i$.

LSTM uses a total of three gates, i.e., input gate, forget gate, and output gate. The input gate determines what information from the current topological embedding and previous temporal embeddings will be cached, or stored for future use, in long term memory. The forget gate decides which information from the long term memory should be maintained or repudiated. The output gate takes the current topological embedding, the previous temporal embedding and the newly computed long term memory to produce the new temporal embedding that will be passed on to the cell in the next time step. The LSTM model can be formulated as follows:
\begin{eqnarray}
&I^{(t)}=\sigma(Z^{(t)}W_{ii}+H^{(t-1)}W_{ih}+b_{i}),\label{eq5}\\[10pt]
&F^{(t)}=\sigma(Z^{(t)}W_{fi}+H^{(t-1)}W_{fh}+b_{f}),\label{eq6}\\[10pt]
&C^{(t)}=F^{(t)}\odot C^{(t-1)}+I^{(t)}\odot tanh(Z^{(t)}W_{ci}+H^{(t-1)}W_{ch}+b_{c}),\label{eq7}\\[10pt]
&O^{(t)}=\sigma(Z^{(t)}W_{oi}+H^{(t-1)}W_{oh}+b_{o}),\label{eq8}\\[10pt]
&H^{(t)}=O^{(t)}\odot tanh(C^{(t)}).\label{eq9}
\end{eqnarray}

In the equations above, $\odot$ denotes element-wise (Hadamard) product. $\sigma$ is an activation function (typically $sigmoid$) and $tanh$ represents the hyperbolic tangent function. $I^{(t)} \in \mathbb{R}^{nl \times D}$, $F^{(t)} \in \mathbb{R}^{nl \times D}$, and $O^{(t)} \in \mathbb{R}^{nl \times D}$ represent input, forget, and output gates for the nodes, respectively. $C^{(t)} \in \mathbb{R}^{nl \times D}$ and $H^{(t)} \in \mathbb{R}^{nl \times D}$ are memory cell and hidden state for the node embeddings, respectively. $W_{(\cdot\cdot)} \in\mathbb{R}^{D \times D}$ and $b_{(\cdot)} \in \mathbb{R}^{1 \times D}$ are weight matrix and bias vector, respectively. $H^{(0)}$ and $C^{(0)}$ can be initialized with zeros or learned from the data \citep{mohajerin2017state}.

GRU is similar to LSTM, but it incorporates two gates, i.e., an update gate and a reset gate. The reset gate determines how much of the previous temporal embedding should be neglected, while the update gate determines the amount of the new input that needs to be passed along to the next state. The GRU model can be formulated as
\begin{eqnarray}
&U^{(t)}=\sigma(Z^{(t)}W_{ui}+H^{(t-1)}W_{uh}+b_{u}),\label{eq10}\\[10pt]
&R^{(t)}=\sigma(Z^{(t)}W_{ri}+H^{(t-1)}W_{rh}+b_{r}),\label{eq11}\\[10pt]
&H^{(t)}=(1-U^{(t)})\odot H^{(t-1)}+U^{(t)}\odot tanh[Z^{(t)}W_{hi}+(R^{(t)}\odot H^{(t-1)})W_{hh}+b_{h}].\label{eq12}
\end{eqnarray}

\begin{figure}[hbt!]
\centering
\begin{subfigure}[b]{0.45\textwidth}
\includegraphics[scale=0.38]{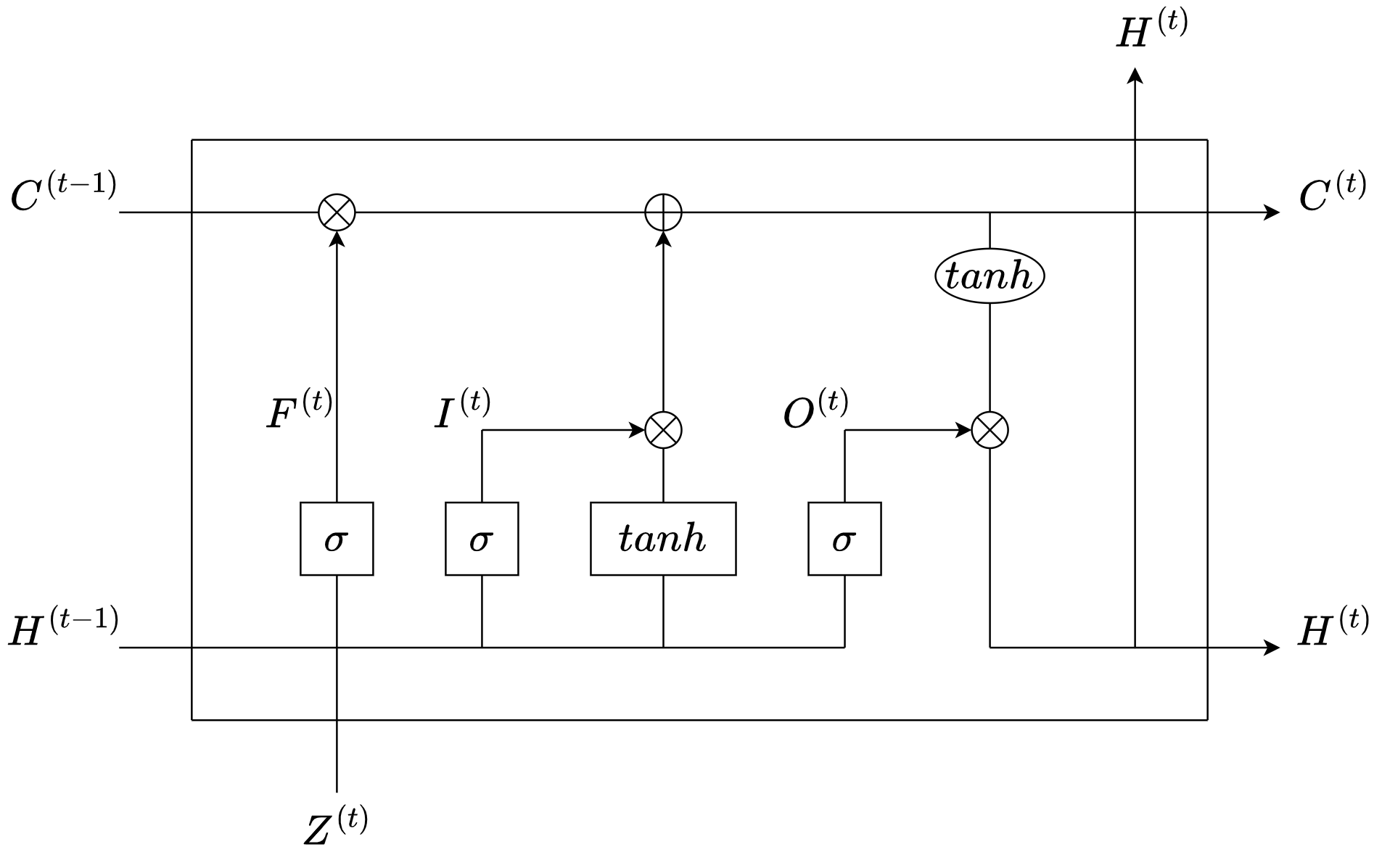}
\caption{LSTM}
\label{fig2a}
\end{subfigure}
\quad
\begin{subfigure}[b]{0.45\textwidth}
\includegraphics[scale=0.38]{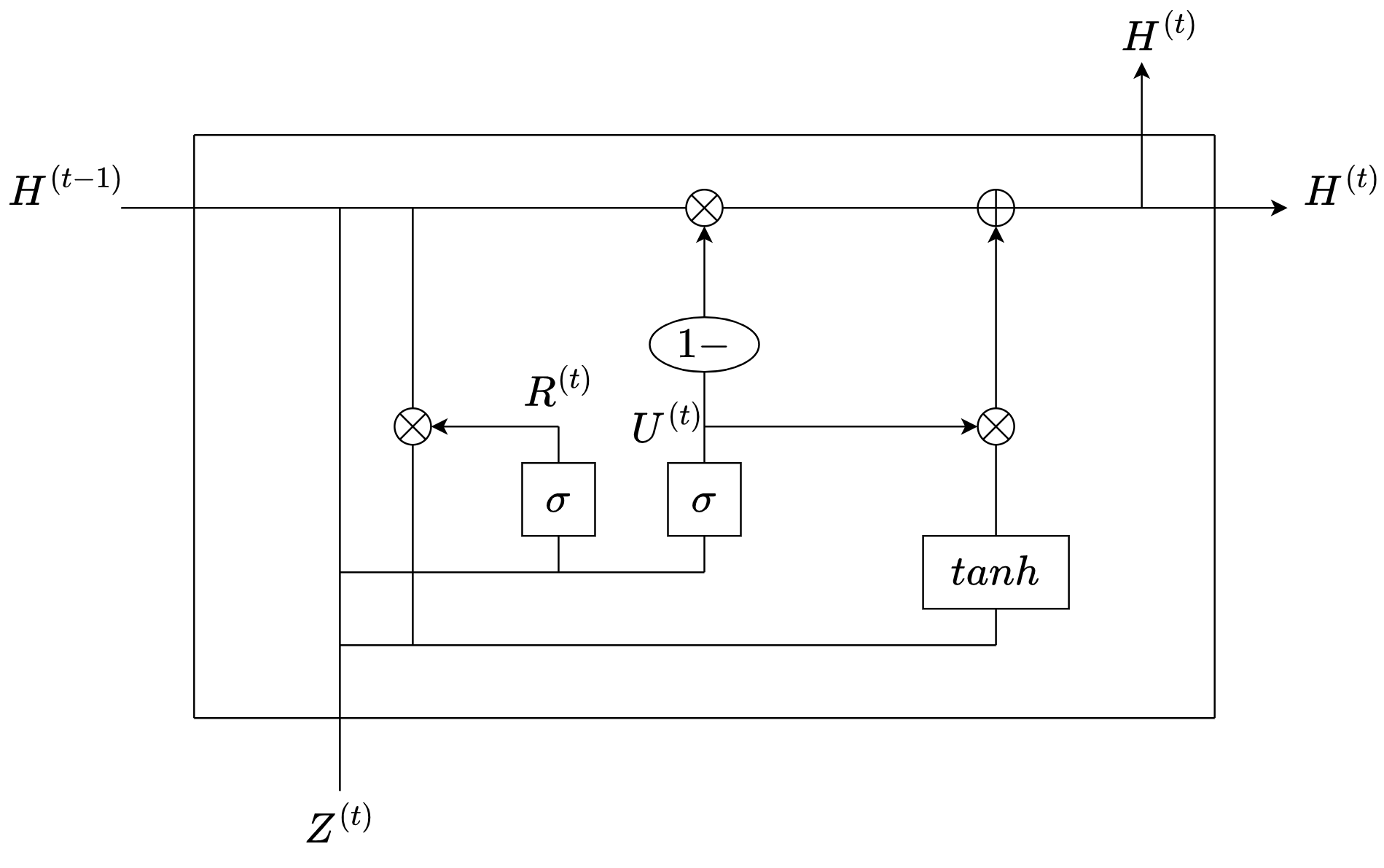}
\caption{GRU}
\label{fig2b}
\end{subfigure}
\caption{The cell structures of RNN models.}
\label{fig2}
\end{figure}

In the equations above, $U^{(t)} \in \mathbb{R}^{nl \times D}$ and $R^{(t)} \in \mathbb{R}^{nl \times D}$ represent update and reset gates for the nodes, respectively. All other definitions are the same as LSTM. Fig.~\ref{fig2} depicts the respective structures of the LSTM and GRU models.

\subsection{GNN-RNN models}
\label{GNN-RNN models}
The GNN-RNN model is one of the proposed models for DYMGNN in this work. Within the GNN-RNN framework, there are two configurations, i.e., GNN-LSTM and GNN-GRU, which can be summarized as
\begin{eqnarray}
&Z^{(t)}=\textrm{GNN}(X^{(t)},A^{(t)}),\label{eq13}\\[10pt]
&H^{(t)},C^{(t)}=\textrm{LSTM}(Z^{(t)},H^{(t-1)},C^{(t-1)}) & \text{for GNN-LSTM},\label{eq14}\\[10pt]
&H^{(t)}=\textrm{GRU}(Z^{(t)},H^{(t-1)}) & \text{for GNN-GRU}.\label{eq15}
\end{eqnarray}

\begin{figure}[hbt!]
\includegraphics[scale=1]{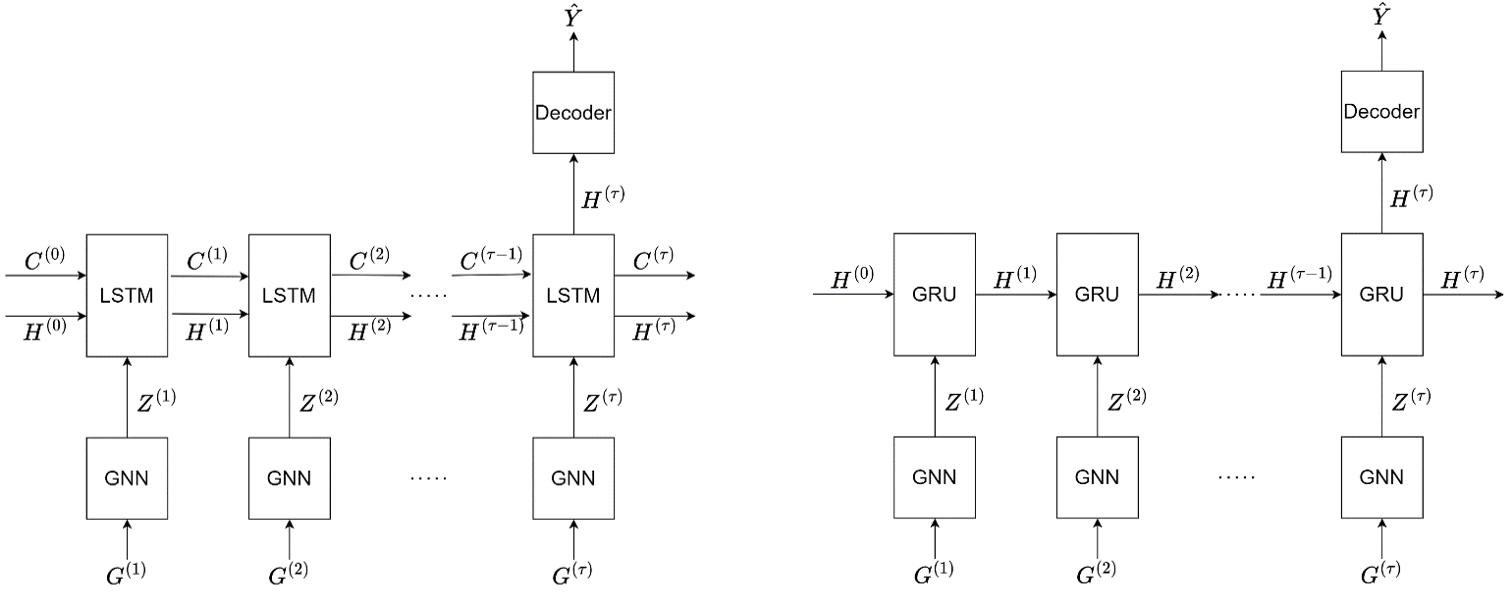}
\centering
\caption{GNN-LSTM (left) and GNN-GRU (right) dynamic models.}
\label{fig3}
\end{figure}

Fig.~\ref{fig3} displays an overview of these models.

These models are capable of capturing the topological and temporal dependencies of snapshots through combining GNN and RNN, whilst the same importance is assigned to each timestamp. The temporal embedding for the nodes is obtained by feeding their sequence of embeddings, produced by the GNN model, to the RNN model.

\subsection{GNN-RNN-ATT models}
\label{GNN-RNN-ATT models}
The GNN-RNN-ATT model is another proposed model for DYMGNN. In GNN-RNN-ATT models, a soft attention mechanism is applied to assign different importance to each timestamp. This approach contrasts with GNN-RNN models which assign equal importance to every timestamp. The use of attention in GNN-RNN-ATT models allows for a more nuanced weighting of temporal information. Our approach for creating a new hidden state for the node embeddings that is more expressive of the global variation trends can be formulated as follows:
\begin{eqnarray}
&s^{(t)}=a_hH^{(t)}W_h,\label{eq16}\\[10pt]
&\beta^{(t)}=\frac{exp(s^{(t)})}{\sum_{k=1}^{\tau}
exp(s^{(k)})},\label{eq17}\\[10pt]
&H_{att}=\sum_{t=1}^{\tau}\beta^{(t)}H^{(t)}.\label{eq18}
\end{eqnarray}

First, the hidden states at different timestamps, $H^{(t)}$, are obtained using GNN and RNN, as discussed in the previous model. Equation \ref{eq16} computes an denormalized attention score for each hidden state, where $a_h \in \mathbb{R}^{1 \times nl}$ and $W_h \in \mathbb{R}^{D \times 1}$ are learnable weight vectors. The normalized attention score for each hidden state is computed using a $softmax$ function as shown in Equation \ref{eq17}. In Equation \eqref{eq18}, $H_{att}$ is calculated by aggregating the hidden states scaled by the normalized attention scores. The main goal of this process is to re-weight the influence of snapshots at different timestamps. Finally, the final output results can be obtained using $H_{att}$ that can describe the global variation information.

\begin{figure}[hbt!]
\includegraphics[scale=1]{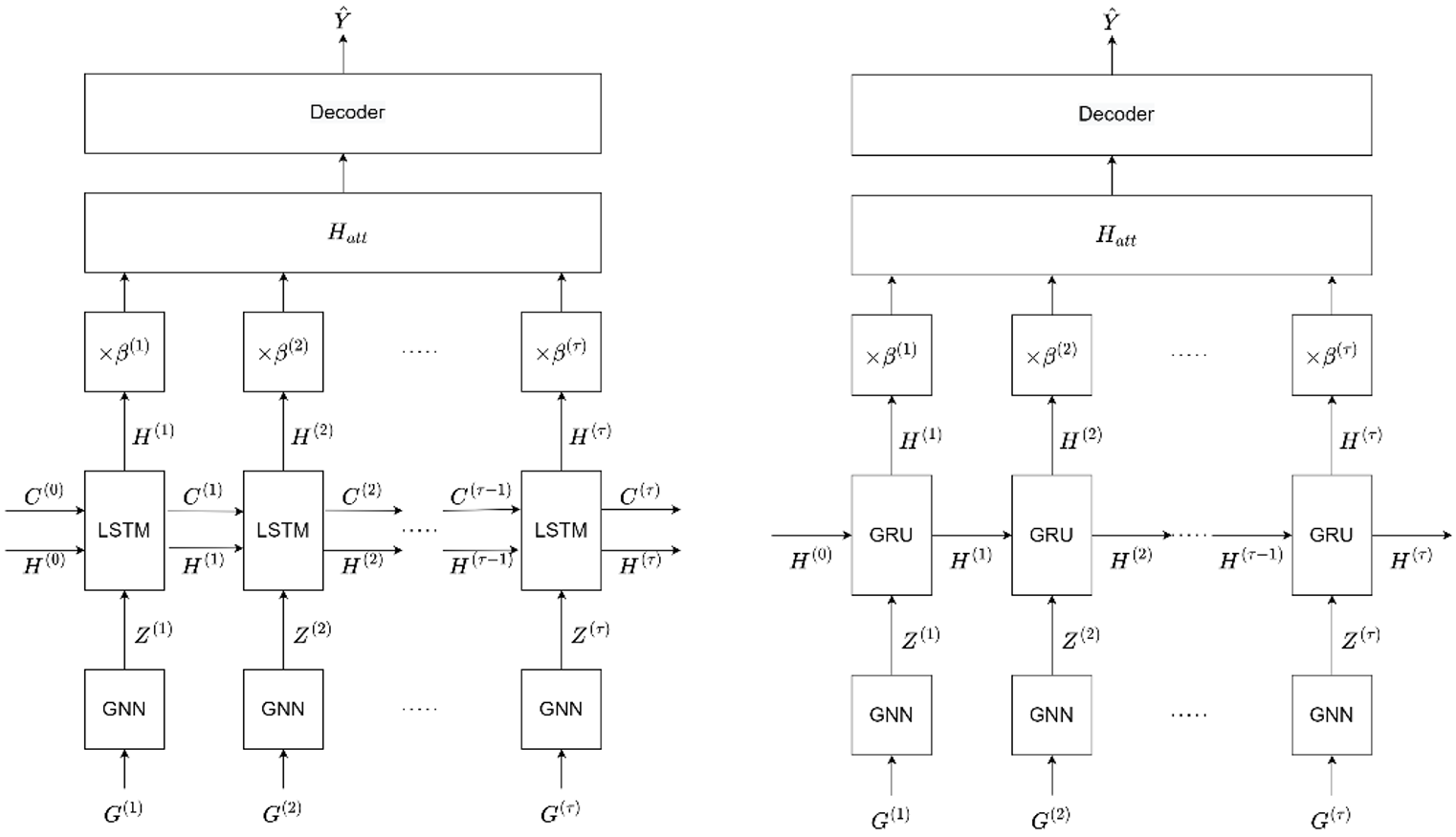}
\centering
\caption{GNN-LSTM-ATT (left) and GNN-GRU-ATT (right) dynamic models. By adding an attention layer to the model, we are able to re-weight the impact of different snapshots.}
\label{fig4}
\end{figure}

Fig.~\ref{fig4} shows two configurations of GNN-RNN-ATT, i.e., GNN-LSTM-ATT and GNN-GRU-ATT.

\subsection{Decoder and loss function}
\label{Decoder and loss function}
\begin{figure}[hbt!]
\includegraphics[scale=0.6]{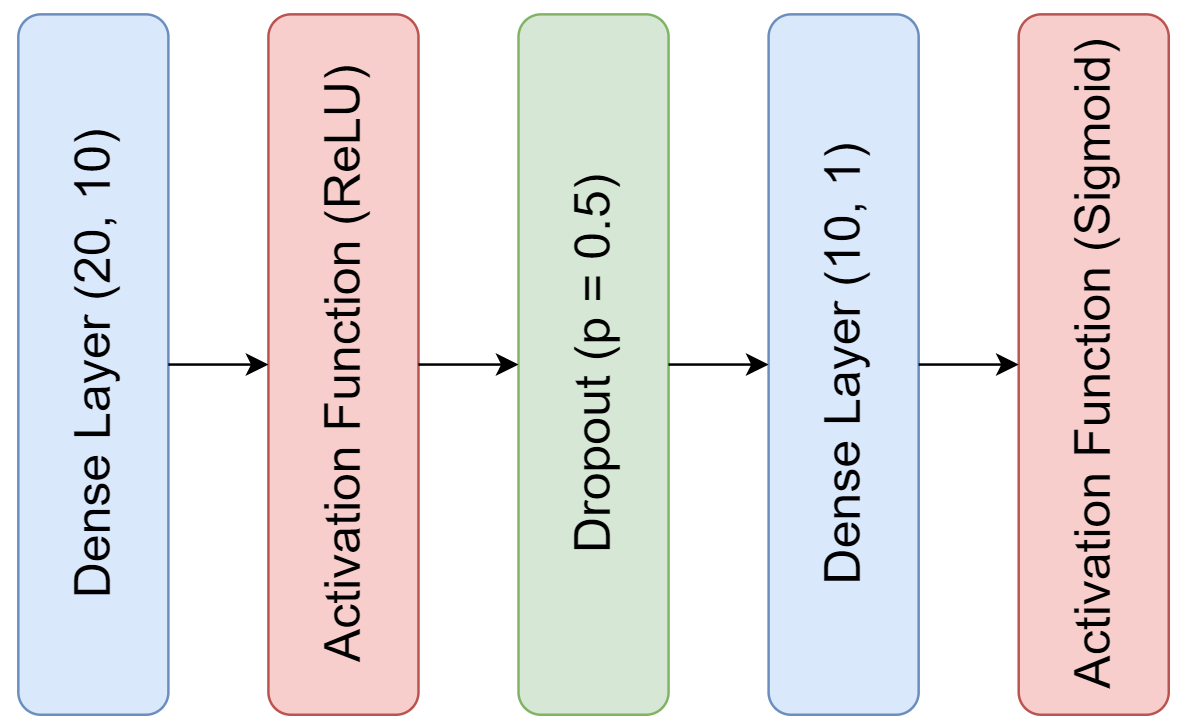}
\centering
\caption{Architecture of the decoder.}
\label{fig5}
\end{figure}

A deep neural network model is typically comprised of an encoder and a decoder. The encoder takes input and produces embeddings, whereas the decoder takes the embeddings and performs the prediction task \citep{GoodBengCour16}. In our specific case, GNN, RNN, and ATT comprise the encoder of the full model, while the decoder is a set of feed-forward neural networks applied to the node embeddings, followed by a series of layers that either apply a chosen activation function for non-linearity or dropout function for regularization. The final output is the model prediction for our binary outcome (here, default Y/N), i.e., whether the node $v_i$ belongs to class $1$ ($Y_i=1$) or $0$ ($Y_i=0$). The decoder outputs a vector $\hat{Y}$ where $\hat{Y}_i$ specifies the probability of a node $v_i$ belonging to class $1$ given the snapshots $[G^{(1)},...,G^{(\tau)}]$. While there is no unique format for the decoder, the architecture used in this work is shown in Fig.~\ref{fig5}.

One of the most important aspects of a deep learning model is its loss function. For our work, we use the well-known binary cross-entropy loss function \citep{gneiting2007strictly} which can be written as
\begin{equation}
\textrm{Loss}=\frac{-1}{n}\sum_{i=1}^{n} \left[Y_i \cdot log(\hat{Y}_i)+(1-Y_i) \cdot log(1-\hat{Y}_i) \right].\label{eq19}
\end{equation}

\section{Experimental setup}
\label{Experimental setup}

\subsection{Dataset}
\label{Dataset}
In this paper, the goal of our models is to predict one-year-ahead loan default based on borrower or loan characteristics. For this purpose, we use the Single-Family Loan-Level (SFLL) dataset provided by the Federal Home Loan Mortgage Corporation (FHLMC), commonly known as Freddie Mac, which contains loan-level data for a sizable share of mortgage loans in the United States \citep{FreddieMac}. Freddie Mac purchases mortgages on the secondary market, pools them, and sells them as a mortgage-backed security to investors on the open market.

The dataset includes information regarding the loan, such as the amount, the interest rate, the insurance percentage, and the provider, as well as information on the borrower, including the borrower’s debt to income ratio and/or unpaid balance, FICO credit score, the geographical area in which they reside, and whether they are a first-time home buyer. It also includes information about the property (type, number of units, etc.). To represent the categorical information, we introduce our own binary features contrasting one category against other categories combined. Numerical node features are normalized using min-max scaling. We clean the data by treating outliers and null values. Specifically, outliers are capped at the 99\textsuperscript{th} percentile and 1\textsuperscript{st} percentile points. There are not many null values, and they are treated with median imputation. Feature descriptions for the data used in model training are given in Table~\ref{tab1}. Most of the features are available at the time of loan application and do not change from one month to another; however, a few features such as `current\textunderscore upb', `if\textunderscore delq\textunderscore sts', `mths\textunderscore remng', and `current\textunderscore int\textunderscore rt' can change from month to month, as they track repayment behaviour over the loan period. More information about the data can be found in Appendix A.

\begin{table}[hbt!]
\small
\caption{Description of the node features.}
\begin{center}
\begin{tabular}{l|l}
\toprule
Feature & Description\\
\midrule
fico & Credit score at the time of acquisition\\
if\textunderscore fthb & Is the borrower a first-time home buyer?\\
mi\textunderscore pct & Mortgage insurance percentage\\
cnt\textunderscore units & Number of units in the property\\
if\textunderscore prim\textunderscore res & Is the property a primary residence?\\
dti & Original debt to income ratio\\
ltv & Original loan to value ratio\\
if\textunderscore corr & Is a correspondent involved in the origination of the mortgage?\\
if\textunderscore sf & Is the property a single family home?\\
if\textunderscore purc & Is the mortgage loan a purchase mortgage?\\
cnt\textunderscore borr & Number of borrowers obligated to repay the mortgage\\
if\textunderscore sc & Does the mortgage exceed conforming loan limit?\\
current\textunderscore upb & Current unpaid principal balance\\
if\textunderscore delq\textunderscore sts & Are there any payment arrears (between 30 and 90 days)?\\
mths\textunderscore remng & Number of remaining months of the mortgage\\
current\textunderscore int\textunderscore rt & Current interest rate\\
default & Being 90 days or more in payment arrears over next 12 months\\
\bottomrule
\end{tabular}
\end{center}
\label{tab1}
\end{table}

\subsection{Dynamic networks}
\label{Dynamic networks}
As we are interested in studying the effect of connections between the borrowers and the evolution of those connections over time, we use the data to create a sequence of dynamic networks following the process in subsection~\ref{Multilayer networks}. In particular, we are interested in predicting one-year-ahead loan default based on application information and six months of borrower's repayment behaviour. We choose a six-month period because six and twelve months are common choices for lookback periods \citep{kennedy2013window}. Furthermore, this paper will later show that extending beyond six months offers minimal additional benefits.

\begin{figure}[hbt!]
\includegraphics[scale=0.8]{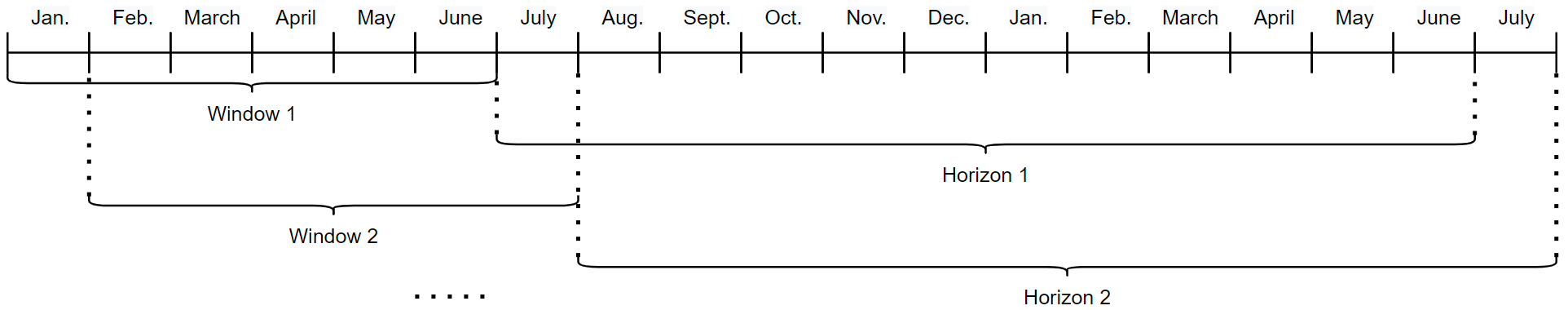}
\centering
\caption{The timeline for windows and their corresponding horizons in model training.}
\label{fig6}
\end{figure}

For this work, loans originated in 2009 and 2010 are used for training and testing. We select these years to ensure sufficient default information is available for reliably comparing the models. It is important to note, however, that loan population and behaviour change over time, and our sample data may reflect some effects of the global financial crisis. Thus, further research could explore the robustness of our proposed methods across different time periods and financial conditions. We use application data and 18 months of behavioural data, from January 2012 to June 2013, for training. We also use application data and six months of behavioural data of a holdout set, from July 2013 to December 2013, for testing. We consider rolling windows, shifting by one month, for training and testing, with each window containing six snapshots $[G^{(1)},...,G^{(6)}]$, and each snapshot corresponding to one month. So, we have 13 windows for training and one window for testing. All snapshots of a specific window have the same set of nodes. However, the node set could be different from one window to another; a loan that has defaulted will remain marked as a defaulter for the observation window but will disappear once the window moves past it. During the training of each window, the goal is to predict default within 12 months following the month of the last snapshot in that window. A one-year horizon is practical for credit management and decision-making, as it balances the need for a sufficiently long period to assess risk while not extending so far that predictions become highly speculative \citep{lopez2000evaluating}. Fig.~\ref{fig6} displays the timeline for the windows and their corresponding horizons in model training.

We can train the models using either single layer or double layer networks, with geographical location of the borrower and the company lending the loan being the connector variables. Borrowers whose zip codes have the same first two digits are assumed to be in the same geographical area. In the case of the double layer network, nodes in one layer are also connected to their twins in the other layer. Some previous studies showed that applying some sort of dropout techniques on graph structures could help increase the expressiveness of the GNN models \citep{shu2022understanding}. Therefore, we decide to randomly select and isolate 50\% of the nodes in each snapshot of a window. In other words, in each snapshot, at least half of the nodes do not have any connections with other nodes.

To gain insight into the size and characteristics of the networks, we provide some descriptions in Table~\ref{tab2}. Letting $G^{(T)}=\bigcup_{t=1}^{6}G^{(t)}$, Table~\ref{tab2} shows the number of nodes and the number of edges for $G^{(T)}$ created from the snapshots in the first window of the training set, as well as the snapshots of the validation and test sets. It is evident that the single layer networks derived from the lending company are denser than those derived from the geographical area. This is because there are fewer lending companies serving as connector variables compared to the number of geographical areas serving as such.

\begin{table}[hbt!]
\footnotesize
\centering
\caption{Network description for $G^{(T)}$.}
\begin{tabular}{l|>{\centering\arraybackslash}p{2cm}>{\centering\arraybackslash}p{2cm}|>{\centering\arraybackslash}p{2cm}>{\centering\arraybackslash}p{2cm}|>{\centering\arraybackslash}p{2cm}>{\centering\arraybackslash}p{2cm}}
\toprule
& \multicolumn{2}{c}{Single Layer:Area} & \multicolumn{2}{c}{Single Layer:Company} & \multicolumn{2}{c}{Double Layer:Area-Company}\\
Set & \multicolumn{1}{c}{\texttt{\#}Nodes} & \multicolumn{1}{c}{\texttt{\#}Edges} & \multicolumn{1}{|c}{\texttt{\#}Nodes} & \multicolumn{1}{c}{\texttt{\#}Edges} & \multicolumn{1}{|c}{\texttt{\#}Nodes} & \multicolumn{1}{c}{\texttt{\#}Edges}\\
\midrule
Training & 148,520 & 16,368,244 & 148,520 & 91,486,176 & 297,040 & 108,151,460\\
Validation & 82,180 & 4,725,842 & 82,180 & 27,735,664 & 164,360 & 32,625,866\\
Test & 96,490 & 6,761,051 & 96,490 & 38,404,277 & 192,980 & 45,358,308\\
\bottomrule
\end{tabular}
\label{tab2}
\end{table}

\subsection{Experiments}
\label{Experiments}
We are interested in comparing the performance of different models on single layer and double layer networks, and in benchmarking them against some baseline methods. The classes are imbalanced in this binary node classification problem, so we use the Area Under the Curve (AUC) and $F_1$ score to assess the performance of each model. The results are presented with 95\% confidence intervals, derived from bootstrap over the test set.

As computational efficiency is another consideration, we also examine the runtime for training the models. In addition, we use the Shapley approach to help us interpret the best performing model and better understand the importance of the different node features. We also look at the attention scores to assess the relative contribution of snapshots at different timestamps.

\section{Results and discussion}
\label{Results and discussion}

\subsection{Baseline methods}
\label{Baseline methods}
We benchmark our proposed model against a selection of GNN-based and non-GNN-based baseline models. Table~\ref{tab3} shows the results for two GNN-based baseline models, i.e., static GCN and static GAT, whereas Table~\ref{tab4} shows the results for three non-GNN-based baseline models, i.e., Logistic Regression (LR), XGBoost (XGB), and a Deep Neural Network (DNN). LR is popular in the commercial and financial sectors due to its straightforwardness and ease of understanding. Meanwhile, XGB has established itself as a powerful technique for both classification and regression tasks involving structured datasets \citep{gunnarsson2021deep}. DNN is fundamental to deep learning and has seen broad usage across a variety of predictive tasks.

For training the GNN-based baseline models, we consider a static network, which is the last snapshot of each window, and values of behavioural features are the mean values of those features across the six snapshots of the respective window. We do not use RNNs for these static models; the decoder and the loss function for these models are the same as those used in the dynamic models.

For non-GNN-based baseline models that rely solely on non-network features, we use a grid search to tune the hyper-parameters for each model using the validation dataset. The LR model is tuned with saga solver and a grid search for the penalty \{L1, L2\}. The XGB model hyper-parameters are tuned with a grid search for the learning rate \{0.001, 0.01, 0.1\}, maximum depth \{2, 3, 4\}, number of estimators \{50, 100, 250, 500\}, and alpha \{0.1,. . . ,0.9\}. The architecture of the DNN is given in Appendix B.

\begin{table}[hbt!]
\footnotesize
\centering
\caption{Performance of the GNN-based baseline models.}
\begin{tabular}{l|cc|cc|cc}
\toprule
& \multicolumn{2}{c}{Single Layer:Area} & \multicolumn{2}{c}{Single Layer:Company} & \multicolumn{2}{c}{Double Layer:Area-Company}\\
Model & \multicolumn{1}{c}{AUC} & \multicolumn{1}{c}{$F_1$} & \multicolumn{1}{|c}{AUC} & \multicolumn{1}{c}{$F_1$} & \multicolumn{1}{|c}{AUC} & \multicolumn{1}{c}{$F_1$}\\
\midrule
Static GCN & $0.701\pm0.014$ & $0.802\pm0.012$ & $0.681\pm0.014$ & $0.798\pm0.013$ & $0.729\pm0.012$ & $0.810\pm0.012$\\
Static GAT & $0.752\pm0.013$ & $0.814\pm0.010$ & $0.746\pm0.011$ & $0.812\pm0.009$ & $0.763\pm0.014$ & $0.817\pm0.012$\\
\bottomrule
\end{tabular}
\label{tab3}
\end{table}

\begin{table}[hbt!]
\small
\centering
\caption{Performance of the non-GNN-based baseline models.}
\begin{tabular}{l|cc}
\toprule
Model & \multicolumn{1}{c}{AUC} & \multicolumn{1}{c}{$F_1$}\\
\midrule
LR & $0.796\pm0.020$ & $0.824\pm0.013$\\
XGB & $0.805\pm0.018$ & $0.837\pm0.012$\\
DNN & $0.803\pm0.016$ & $0.833\pm0.014$\\
\bottomrule
\end{tabular}
\label{tab4}
\end{table}

In Table~\ref{tab3}, we can see that the Static GAT performs better than the Static GCN, on both single layer and double layer networks. The difference in performance between them is considerable, and could be due to the different way in which GAT and GCN aggregate information from the one-hop neighbourhood. Among the non-GNN models (see Table~\ref{tab4}), XGB appears to have better performance compared to LR and DNN; however, the differences are fairly small. XGB outperforms LR, suggesting that XGB can capture non-linear relationships better than LR does. It is also not unexpected to see that DNN does not outperform XGB, as this might be the case where the structured data is not very complex or does not contain many features \citep{borisov2022deep, gunnarsson2021deep}. Comparing Table~\ref{tab4} against Table~\ref{tab3}, it is observed that the performance of each non-GNN-based baseline model surpasses that of the best performing GNN-based baseline model, i.e., Static GAT, on both single layer and double layer networks. This observation is important as it indicates that more complex models do not always yield better performance.

\subsection{Performance of the dynamic models}
\label{Performance of the dynamic models}
Table~\ref{tab5} and Table~\ref{tab6} show the performance of the different models on our two single layer networks, while Table~\ref{tab7} shows the performance on the double layer network. Out-of-sample performance is again measured in terms of AUC and $F_1$ score on a test set. The best performing model in each group is highlighted in bold.

\begin{table}[hbt!]
\small
\centering
\caption{Performance of the dynamic models on the single layer network derived from the geographical area.}
\begin{tabular}{ll|cc|cc}
\toprule
\multicolumn{2}{c}{Model}& \multicolumn{2}{c}{AUC} & \multicolumn{2}{c}{$F_1$}\\
Topological&Temporal
 & \multicolumn{1}{|c}{Without ATT} & \multicolumn{1}{c}{With ATT} & \multicolumn{1}{|c}{Without ATT} & \multicolumn{1}{c}{With ATT}\\
\midrule
\multirow{2}{*}{GCN}&LSTM & $0.804\pm0.011$ & $0.807\pm0.012$ & $0.841\pm0.008$ & $0.847\pm0.009$\\
&GRU & $0.775\pm0.013$ & $0.780\pm0.011$ & $0.825\pm0.006$ & $0.829\pm0.008$\\
\multirow{2}{*}{GAT}&LSTM & $0.806\pm0.009$ & $\mathbf{0.810\pm0.012}$ & $0.842\pm0.005$ & $\mathbf{0.849\pm0.007}$\\
&GRU & $0.793\pm0.005$ & $0.802\pm0.014$ & $0.833\pm0.004$ & $0.840\pm0.006$\\
\bottomrule
\end{tabular}
\label{tab5}
\end{table}

\begin{table}[hbt!]
\small
\centering
\caption{Performance of the dynamic models on the single layer network derived from the lending company.}
\begin{tabular}{ll|cc|cc}
\toprule
\multicolumn{2}{c}{Model}& \multicolumn{2}{c}{AUC} & \multicolumn{2}{c}{$F_1$}\\
Topological&Temporal & \multicolumn{1}{|c}{Without ATT} & \multicolumn{1}{c}{With ATT} & \multicolumn{1}{|c}{Without ATT} & \multicolumn{1}{c}{With ATT}\\
\midrule
\multirow{2}{*}{GCN}&LSTM & $0.802\pm0.012$ & $0.804\pm0.012$ & $0.840\pm0.009$ & $0.843\pm0.007$\\
&GRU & $0.769\pm0.013$ & $0.774\pm0.014$ & $0.818\pm0.006$ & $0.823\pm0.006$\\
\multirow{2}{*}{GAT}&LSTM & $0.805\pm0.012$ & $\mathbf{0.808\pm0.010}$ & $0.840\pm0.009$ & $\mathbf{0.846\pm0.008}$\\
&GRU & $0.786\pm0.007$ & $0.795\pm0.014$ & $0.832\pm0.004$ & $0.835\pm0.006$\\
\bottomrule
\end{tabular}
\label{tab6}
\end{table}

Table~\ref{tab5} and Table~\ref{tab6} show that, for each of the single layer networks, the GAT-LSTM-ATT model produces the highest AUC and $F_1$ score, while GCN-GRU gives the poorest results. This could be due to the fact that GAT assigns different importance to each edge, and we know that some connections could be more informative than others. Also, the complex structure of LSTM appears to make it the preferred RNN for this problem. Another key observation is that models enhanced with the attention mechanism consistently show better performance compared to those without attention.

\begin{table}[hbt!]
\small
\centering
\caption{Performance of the dynamic models on double layer network created with both geographical area and lending company.}
\begin{tabular}{ll|cc|cc}
\toprule
\multicolumn{2}{c}{Model}& \multicolumn{2}{c}{AUC} & \multicolumn{2}{c}{$F_1$}\\
Topological&Temporal & \multicolumn{1}{|c}{Without ATT} & \multicolumn{1}{c}{With ATT} & \multicolumn{1}{|c}{Without ATT} & \multicolumn{1}{c}{With ATT}\\
\midrule
\multirow{2}{*}{GCN}&LSTM & $0.806\pm0.010$ & $0.810\pm0.009$ & $0.845\pm0.004$ & $0.848\pm0.006$\\
&GRU & $0.789\pm0.010$ & $0.793\pm0.011$ & $0.833\pm0.005$ & $0.835\pm0.009$\\
\multirow{2}{*}{GAT}&LSTM & $0.807\pm0.008$ & $\mathbf{0.812\pm0.008}$ & $0.847\pm0.005$ & $\mathbf{0.851\pm0.007}$\\
&GRU & $0.800\pm0.004$ & $0.804\pm0.006$ & $0.839\pm0.003$ & $0.843\pm0.008$\\
\bottomrule
\end{tabular}
\label{tab7}
\end{table}

As for the double layer network, we can see from Table~\ref{tab7} that, similarly to what was observed for the single layer networks, the GAT-LSTM-ATT model again shows the best performance. The results for the double layer network, however, tend to outperform the single layer ones, which is intuitive as the double layer network is able to consider connections of either type. It is also noticeable that the results obtained from the double layer network have shorter confidence intervals, suggesting greater robustness in these results. Importantly, the best performing dynamic modelling approach, i.e., GAT-LSTM-ATT, performs better on average than the baseline methods presented in the previous subsection. For example, the GAT-LSTM-ATT model for the double-layer network produces AUC and $F_1$ score of 0.812 and 0.851, respectively, compared to 0.805 and 0.837 for the best baseline model, i.e., XGB (see Table~\ref{tab4}). This translates to a 0.87\% gain in AUC and a 1.67\% gain in $F_1$ score. Although these numerical gains might seem modest, even a 1\% improvement can yield significant financial benefits for some businesses. Interestingly, even when applied to either of the single layer networks, the GAT-LSTM-ATT still tends to perform well compared to the baseline models. This demonstrates that DYMGNN offers an advantage over conventional methods by capturing a richer set of information, thus providing a more comprehensive and realistic picture of a borrower's default probability. Hence, incorporating dynamic network information is able to provide additional information over simply using local features and/or static networks. The most pronounced difference lies between our dynamic model and the static network-based models, demonstrating the importance of capturing network changes over a sufficiently long time window.

\subsection{Runtime analysis}
\label{Runtime analysis}
Computational complexity of training machine learning models considers two core aspects: time complexity and space complexity. Time complexity relates to the time it takes to train a model and how this is affected by problem size, whereas space refers to how much space a model uses (memory footprint).

As time complexity is a potential consideration in our work, we report the runtimes for the dynamic models in Table~\ref{tab8} and Table~\ref{tab9}. The runtimes for the GNN-RNN models and GNN-RNN-ATT models are presented in separate tables as those models' architectures differ from each other. The hyperparameters and resources used for the computations can be found in Appendix C. Note, to allow for easier comparison, the runtimes in each table are also normalized with respect to the lowest number in that table.

\begin{table}[hbt!]
\scriptsize
\centering
\caption{Runtime for training the GNN-RNN models (seconds).}
\begin{tabular}{ll|cc|cc|cc}
\toprule
\multicolumn{2}{c}{Model}& \multicolumn{2}{c}{Single Layer:Area} & \multicolumn{2}{c}{Single Layer:Company} & \multicolumn{2}{c}{Double Layer:Area-Company}\\
Topological&Temporal & \multicolumn{1}{|c}{Non-normalized} & \multicolumn{1}{c}{Normalized} & \multicolumn{1}{|c}{Non-normalized} & \multicolumn{1}{c}{Normalized} & \multicolumn{1}{|c}{Non-normalized} & \multicolumn{1}{c}{Normalized}\\
\midrule
\multirow{2}{*}{GCN}&LSTM & 1,690 & 1.18 & 7,090 & 4.93 & 8,397 & 5.84\\
&GRU & 1,437 & 1.00 & 6,109 & 4.25 & 7,221 & 5.03\\
\multirow{2}{*}{GAT}&LSTM & 2,571 & 1.79 & 10,148 & 7.06 & 12,171 & 8.47\\
&GRU & 2,081 & 1.45 & 8,390 & 5.84 & 10,019 & 6.97\\
\bottomrule
\end{tabular}
\label{tab8}
\end{table}

\begin{table}[hbt!]
\scriptsize
\centering
\caption{Runtime for training the GNN-RNN-ATT models (seconds).}
\begin{tabular}{ll|cc|cc|cc}
\toprule
\multicolumn{2}{c}{Model}& \multicolumn{2}{c}{Single Layer:Area} & \multicolumn{2}{c}{Single Layer:Company} & \multicolumn{2}{c}{Double Layer:Area-Company}\\
Topological&Temporal & \multicolumn{1}{|c}{Non-normalized} & \multicolumn{1}{c}{Normalized} & \multicolumn{1}{|c}{Non-normalized} & \multicolumn{1}{c}{Normalized} & \multicolumn{1}{|c}{Non-normalized} & \multicolumn{1}{c}{Normalized}\\
\midrule
\multirow{2}{*}{GCN}&LSTM & 1,700 & 1.16 & 7,104 & 4.86 & 8,481 & 5.80\\
&GRU & 1,463 & 1.00 & 6,190 & 4.23 & 7,225 & 4.94\\
\multirow{2}{*}{GAT}&LSTM & 2,597 & 1.78 & 10,151 & 6.94 & 12,120 & 8.28\\
&GRU & 2,114 & 1.44 & 8,413 & 5.75 & 10,054 & 6.87\\
\bottomrule
\end{tabular}
\label{tab9}
\end{table}

From the tables, we can see that training a model on a single layer network derived from the lending company takes longer than training a model on a single layer network created based on geographical area. This is not unexpected as the former network contains a higher number of connections between the nodes compared to the latter. GAT-LSTM and GAT-LSTM-ATT have the highest training runtimes among the GNN-RNN and GNN-RNN-ATT models, respectively, whereas GCN-GRU and GCN-GRU-ATT have the lowest runtimes. GNN-RNN-ATT models normally have higher runtimes compared to the GNN-RNN models, owing to the complexity added to those models by the attention mechanism. It is worth noting that the runtime for the XGB model is only 151 seconds, highlighting the greater complexity of our proposed models compared to traditional non-network classifiers.

\subsection{Interpretability of the architecture}
\label{Interpretability of the architecture}
\begin{figure}[hbt!]
\centering
\begin{subfigure}[b]{0.50\textwidth}
\includegraphics[scale=0.80]{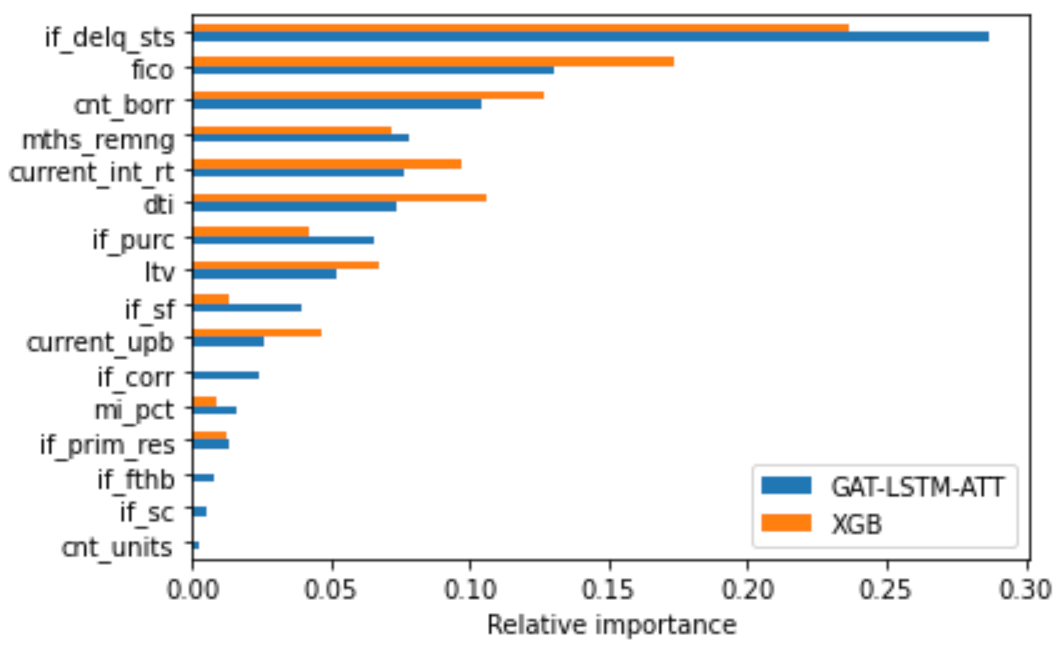}
\caption{Relative importance for GAT-LSTM-ATT and XGB.}
\label{fig7a}
\end{subfigure}
\quad
\begin{subfigure}[b]{0.40\textwidth}
\includegraphics[scale=0.45]{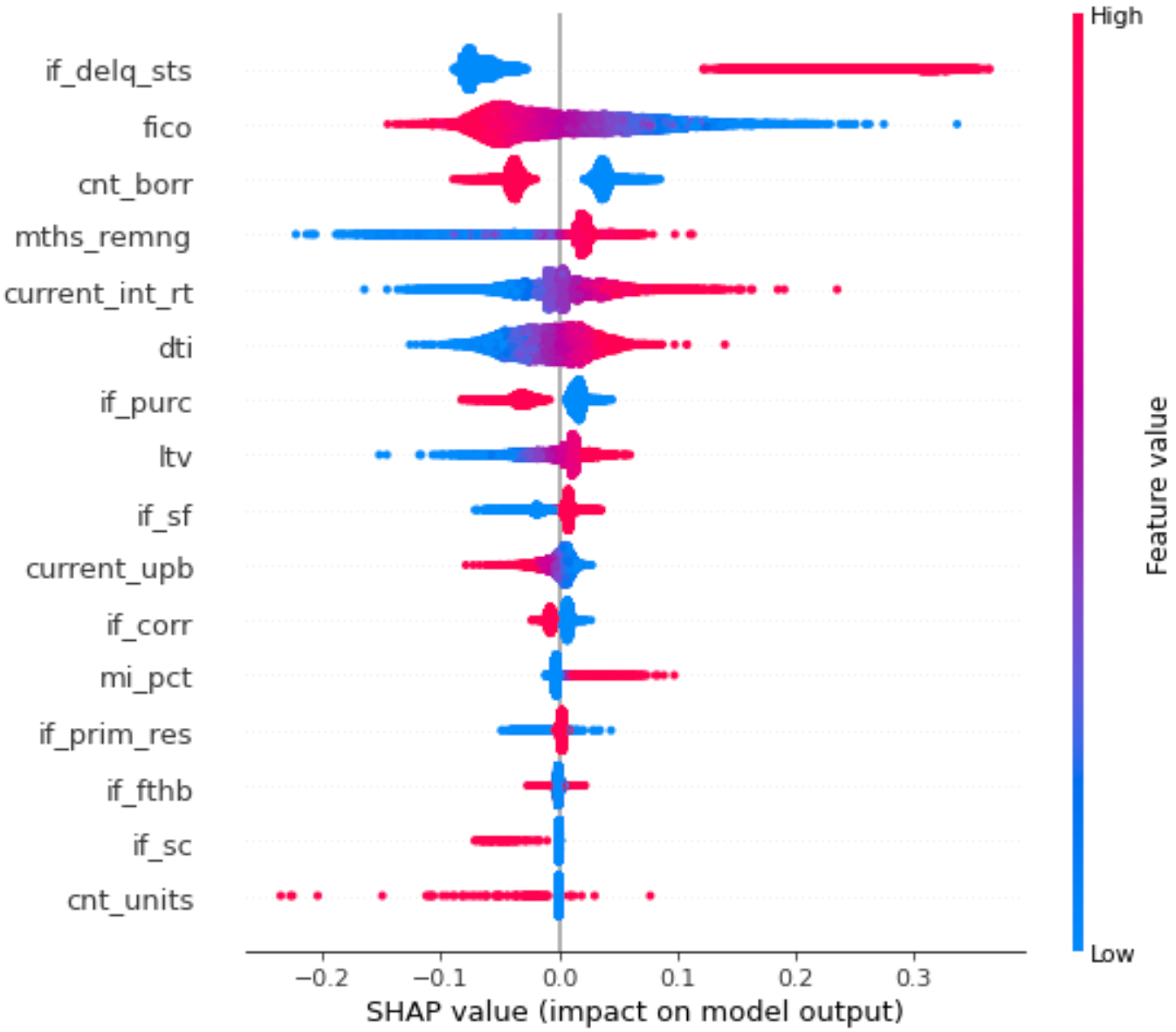}
\caption{Shapley values for GAT-LSTM-ATT.}
\label{fig7b}
\end{subfigure}
\caption{Summary of node feature importance.}
\label{fig7}
\end{figure}

Having established that the best results can be obtained by applying the GAT-LSTM-ATT model to the double layer network, in this section, we employ the Shapley approach \citep{lundberg2017unified} to better understand this model. Using this method, we can establish each node feature’s relative importance and quantify its contribution to the model output. Fig.~\ref{fig7a} displays the relative importance of node features for the best performing proposed model, i.e., GAT-LSTM-ATT, and the best performing baseline model, i.e., XGB. Fig.~\ref{fig7b} displays an information-dense summary of how the node features for the best performing proposed model impact its output.

As seen in Fig.~\ref{fig7a}, the presence of overdue payments holds the most significant relative importance compared to other features, for both GAT-LSTM-ATT and XGB. Overdue payments are a strong indicator of a borrower's financial health; similarly, while timely payments generally suggest good financial management, payment arrears can signal financial distress. The FICO credit score has the second highest relative contribution among the features. This is intuitive as this feature summarizes a lot of information about the payment history and financial behaviour of the borrower. For both models, the number of borrowers ranks as the third most crucial feature. For the GAT-LSTM-ATT model, the number of remaining months holds the fourth position in terms of importance, whereas for the XGB model, the debt to income ratio claims the fourth spot. Notably, the disparity in the significance attributed to features by the two models is more pronounced for the top two features.

Fig.~\ref{fig7b} shows that payment arrears are highly indicative of default risk. It can also be viewed that borrowers with high credit scores are less likely to default, according to the model, while borrowers with low credit scores are more prone to be classified as defaulters. High values of the number of borrowers are associated with lower default risk, while low values are associated with higher default risk. Higher (lower) number of remaining months is linked with higher (lower) default risk.

\begin{figure}[hbt!]
\centering
\begin{subfigure}[b]{0.5\textwidth}
\includegraphics[scale=0.6]{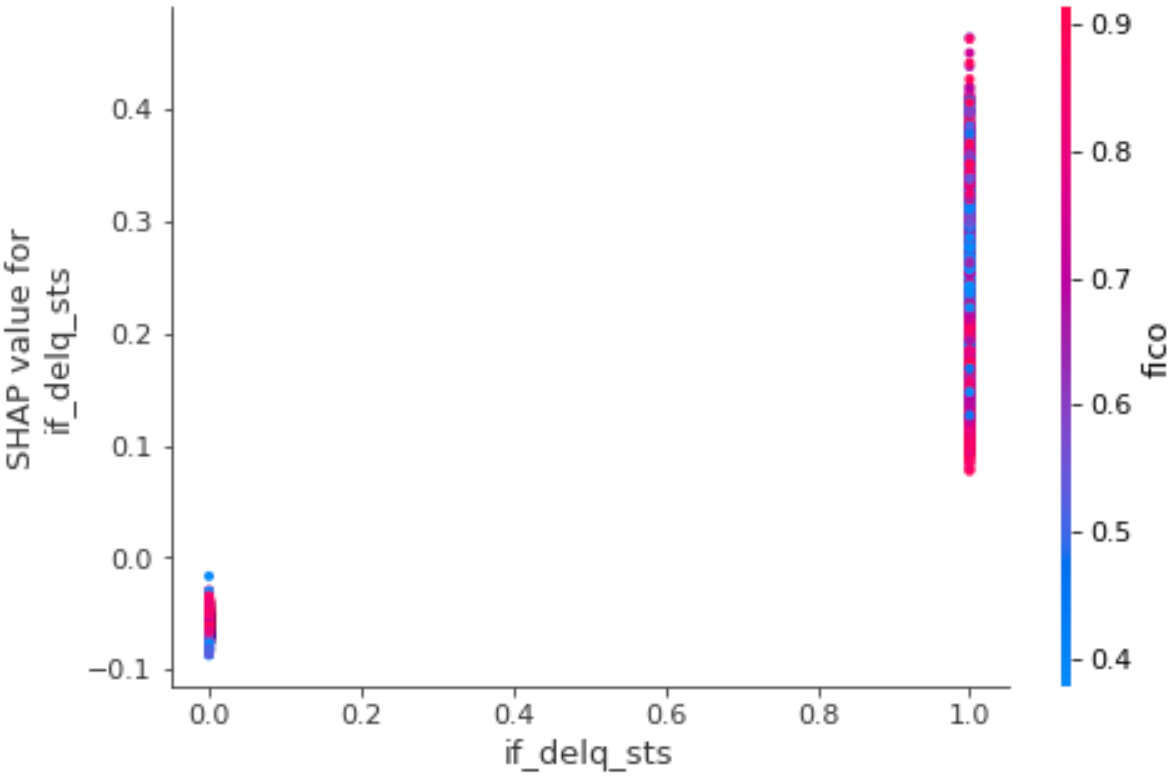}
\caption{if\_delq\_sts}
\label{fig8a}
\end{subfigure}
\quad
\begin{subfigure}[b]{0.42\textwidth}
\includegraphics[scale=0.6]{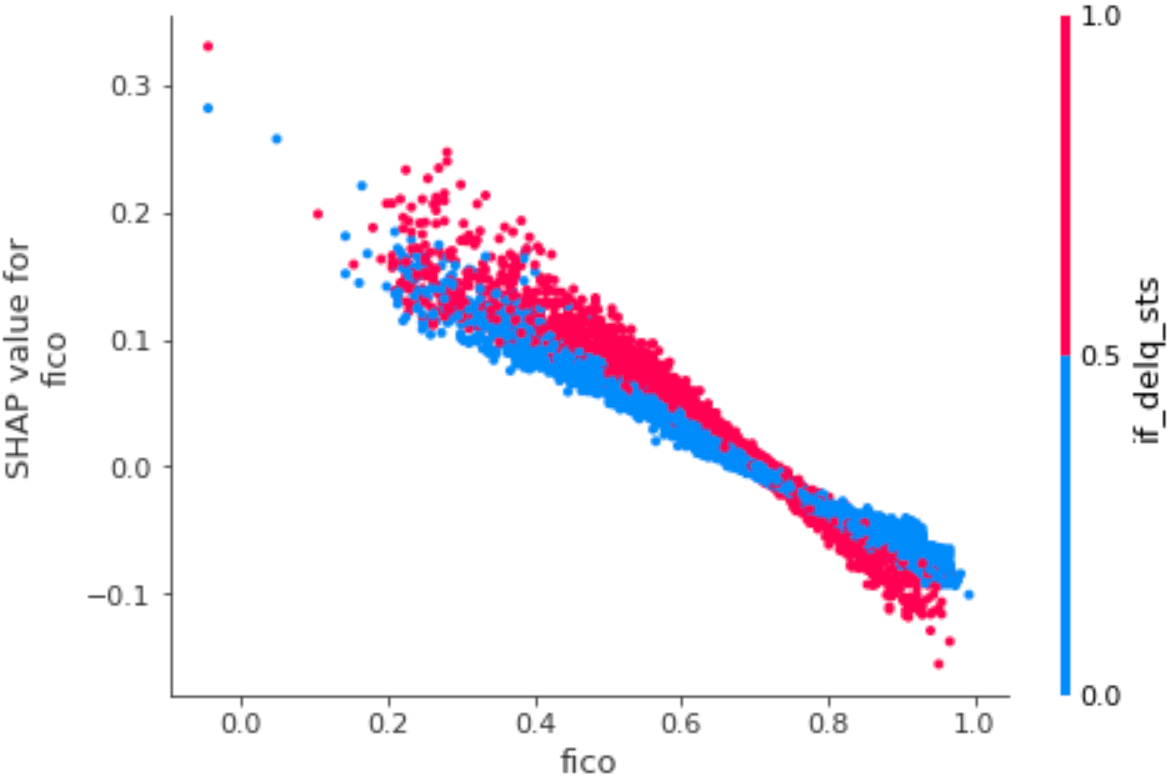}
\caption{fico}
\label{fig8b}
\end{subfigure}
\newline
\newline
\begin{subfigure}[b]{0.5\textwidth}
\includegraphics[scale=0.6]{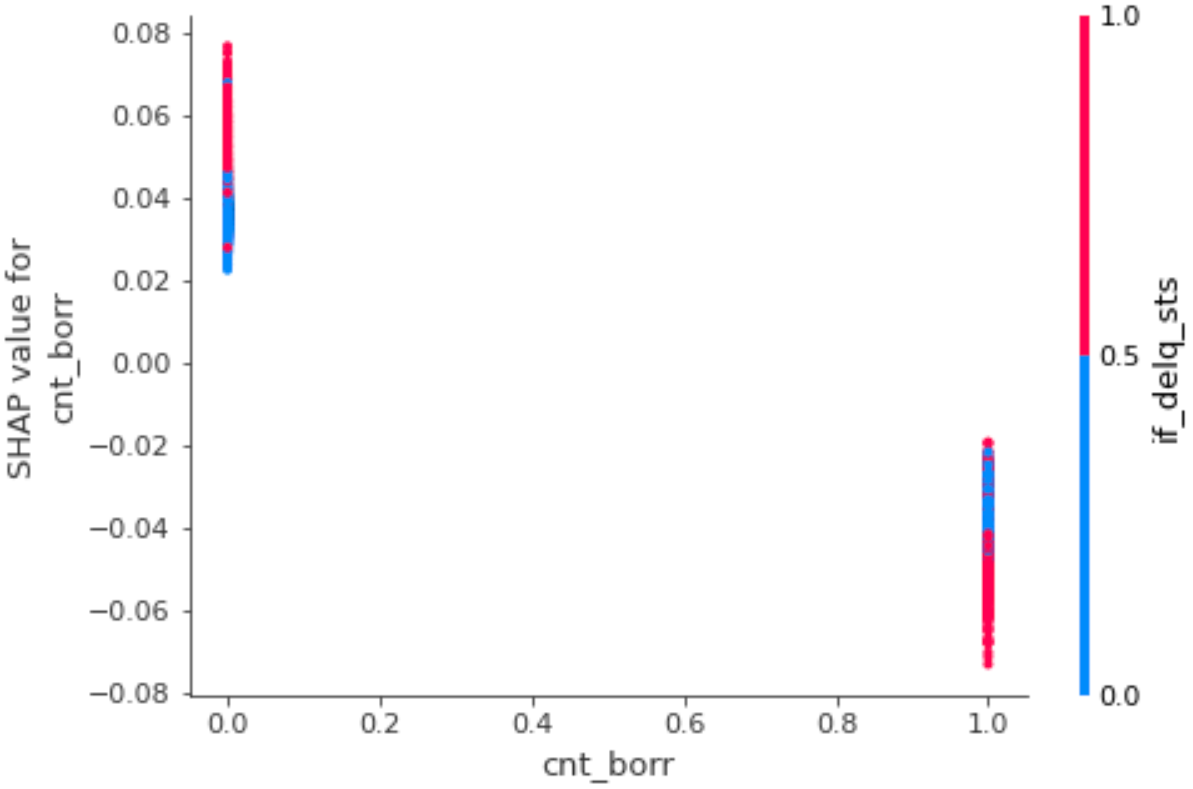}
\caption{cnt\_borr}
\label{fig8c}
\end{subfigure}
\quad
\begin{subfigure}[b]{0.45\textwidth}
\includegraphics[scale=0.6]{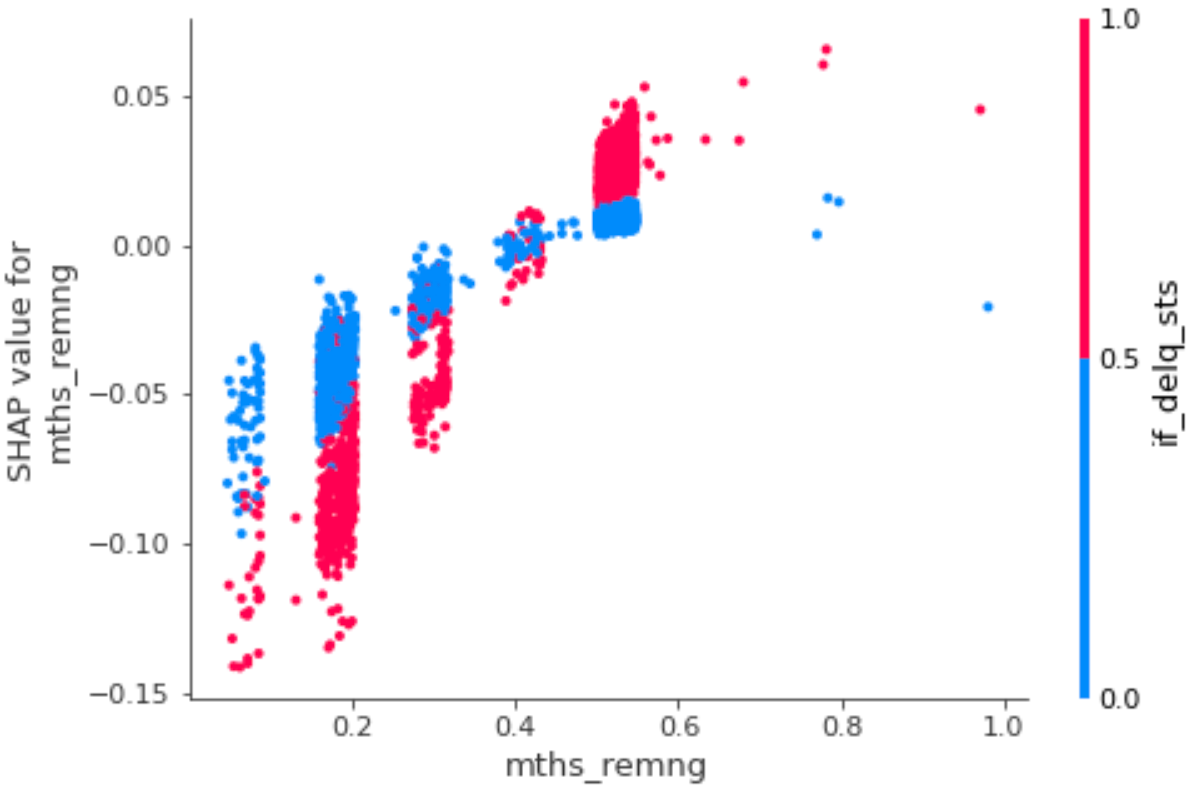}
\caption{mths\_remng}
\label{fig8d}
\end{subfigure}
\caption{Dependency plots of the four most important features for GAT-LSTM-ATT. The colour shows the value of the closest feature by correlation.}
\label{fig8}
\end{figure}

Fig.~\ref{fig8} displays the dependency plots of the four most important features. Note, the feature values are scaled to be between $0$ and $1$ using min-max scaling. Fig.~\ref{fig8a} illustrates that borrowers who consistently meet their payment deadlines tend to have high credit scores and are less prone to default. On the other hand, those who experience delays in making payments are more commonly associated with the cohort of defaulters. Fig.~\ref{fig8b} demonstrates that the credit score, by and large, has a linear impact, with higher scores signalling lower risk of default. Lower values of credit score are much more informative than the higher values. Additionally, the model reveals some intriguing interaction effects. The relative impact of credit score on the default risk is more pronounced in the case of borrowers who have a history of late payments. This indicates that while a low credit score is generally a good indicator of high default risk, its predictive influence increases for those who do not consistently make timely payments. Fig.~\ref{fig8c} suggests that cases involving fewer borrowers are more likely to default on their loans. This might be attributed to various factors, such as limited financial resources, reduced collective responsibility, or lesser peer pressure to maintain creditworthiness among a smaller group. Fig.~\ref{fig8d} indicates that the number of remaining months displays a nearly linear trend, with lower numbers pointing to safer cases. This may stem from the increased uncertainty associated with longer durations (as opposed to shorter durations which signal an approaching end to the financial commitment) or from survival bias. Additionally, the figure points out that this feature's effect on default risk is more significant for borrowers with a pattern of delayed payments.

\begin{figure}[hbt!]
\includegraphics[scale=0.8]{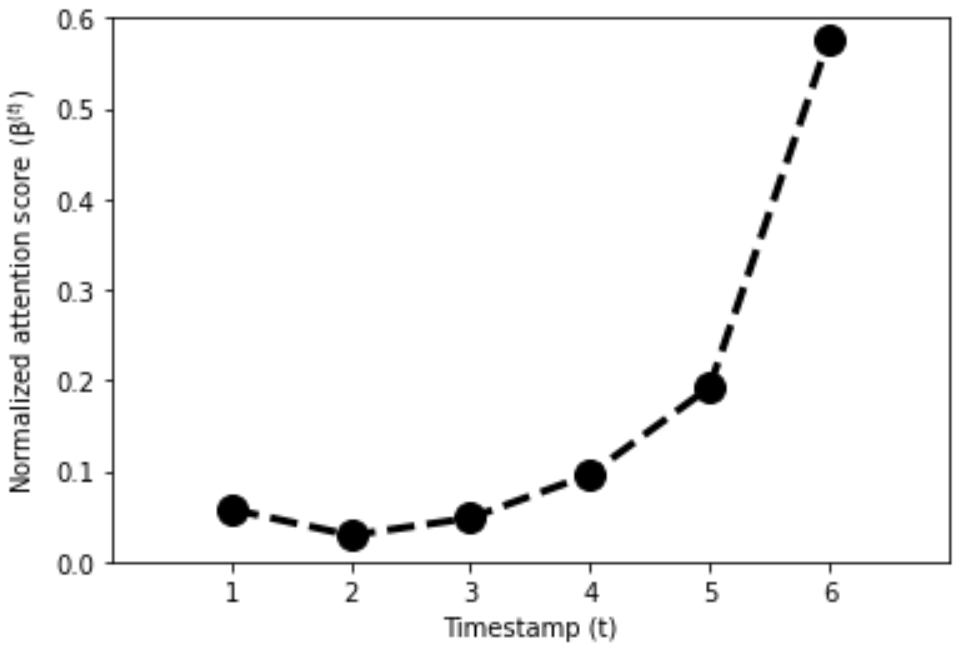}
\centering
\caption{Variation of the normalized attention scores.}
\label{fig9}
\end{figure}

We also aim to analyse the normalized attention scores from the GAT-LSTM-ATT model to determine the relative importance of each timestamp. Fig.~\ref{fig9} illustrates how these scores vary over time.

The figure shows that for the first few timestamps, the attention score is relatively stable and low, remaining close to 0.1. This indicates a minimal level of attention or importance being assigned during these early timestamps. However, as time progresses, particularly after timestamp 3, there is a noticeable upward trend in the attention score. This increase becomes more pronounced between timestamps 4 and 6, where the attention score rises sharply, peaking just below 0.6 at timestamp 6. This pattern suggests that as time progresses, the snapshots grow in importance. The reason for this progressive increase in attention could be that the most recent information holds greater value. Additionally, it can be inferred that longer lookback periods are unlikely to add anything extra, since the bulk of attention is allocated within a relatively brief period.

\section{Conclusions}
\label{Conclusions}
This study introduced an innovative approach to credit risk assessment through the use of dynamic graph neural networks. We found that this technique outperforms traditional models commonly applied in the sector when tested against US mortgage data. By harnessing the capabilities of both GNN and RNN, our method successfully captures the evolving connections between individual loans. We engineered this methodology to exploit the potential of multilayer networks, rather than the common single layer ones. The findings suggested that models incorporating double layer networks with a customized attention mechanism show enhanced predictive capability.

The evaluation of these models was conducted using a dataset from the mortgage lending domain. In our experiments, we constructed single and double layer networks using the borrower's geographical location and the lending company as the connector variables. Through rigorous testing of various models, we established that the GAT-LSTM-ATT model exhibits the best performance among all configurations of DYMGNN, and other baseline models, both GNN-based and non-GNN-based. Furthermore, this model is able to capture a richer set of information, thereby providing more realistic insights into a borrower's probability of default.

When comparing training times, it became apparent that models employing the attention mechanism exhibit greater complexity and require more extensive training times, yet the runtime remains within acceptable limits. As in any operational research area, explainability is important to consider \citep{de2023explainable}. Therefore, we applied the Shapley approach to decode the model's inner workings, assessing the impact of each node feature on the final output. This analysis revealed the differences in the importance of node features between a baseline model and our DYMGNN model, with particular focus on the four most pivotal features. Additionally, we investigated the relative importance of snapshots at different timestamps by analysing the attention scores associated with each. The results confirmed that the most recent snapshots play a crucial role in influencing the model's output.

Future research could explore wider networks by incorporating additional layers to map more complex inter-individual connections. The method could also be extended by further considering distance information and assigning different weights to the network edges based on geographical proximity between the centroids of neighbouring zip code areas. It might also be beneficial to vary the number of snapshots by adjusting the window length for generating dynamic networks. The exploration of other GNNs and RNNs not considered in this study presents another promising direction. Moreover, gaining a deeper understanding of how different network connections influence default risk could offer valuable insights into credit risk modelling.

\section*{Acknowledgements}
\label{Acknowledgements}
The first author acknowledges the support of the Natural Sciences and Engineering Research Council (NSERC) of Canada through the Canada Graduate Scholarships – Doctoral (CGS D) program. The second and fourth authors acknowledge the support of the Economic and Social Research Council (ESRC) [grant number ES/P000673/1]. The third author acknowledges the support of the Icelandic Research Fund (IRF) [grant number 228511-051]. The last author acknowledges the support of the NSERC [discovery grant RGPIN-2020-07114]. This research was undertaken, in part, thanks to funding from the Canada Research Chairs program [CRC-2018-00082]. This work was enabled in part by support provided by Compute Ontario (\url{computeontario.ca}), Calcul Québec (\url{calculquebec.ca}), and the Digital Research Alliance of Canada (\url{alliancecan.ca}).

\bibliographystyle{model5-names.bst}
{\bibliography{DYMGNN.bib}}

\begin{appendices}
\counterwithin{table}{section}
\counterwithin{figure}{section}
\renewcommand\thetable{\Alph{section}.\arabic{table}}
\renewcommand\thefigure{\Alph{section}.\arabic{figure}}

\newpage

\section{Statistics of the node features}
\label{Statistics of the node features}

\begin{table}[hbt!]
\footnotesize
\caption{Descriptive statistics of the non-binary node features. For each loan's behavioural features (`current\textunderscore upb', `mths\textunderscore remng', and `current\textunderscore int\textunderscore rt), the maximum values over all monthly snapshots are considered.}
\begin{center}
\begin{tabular}{l|cccc}
\toprule
Feature & Mean & Std. Dev. & Min. & Max.\\
\midrule
fico & 752.76 & 44.75 & 565 & 832\\
mi\textunderscore pct & 2.40 & 7.40 & 0 & 35\\
cnt\textunderscore units & 1.02 & 0.17 & 1 & 4\\
dti & 33.61 & 11.15 & 1 & 65\\
ltv & 69.30 & 16.07 & 7 & 97\\
cnt\textunderscore borr & 1.50 & 0.50 & 1 & 2\\
current\textunderscore upb & 173,036.60 & 97,258.30 & 13,829.33 & 716,617.50\\
mths\textunderscore remng & 304.58 & 65.55 & 73 & 574\\
current\textunderscore int\textunderscore rt & 4.88 & 0.45 & 3.25 & 7.25\\
\bottomrule
\end{tabular}
\end{center}
\end{table}

\begin{table}[hbt!]
\footnotesize
\caption{Frequency of the binary node features. For each loan, the maximum value of `if\textunderscore delq\textunderscore sts' over all monthly snapshots is considered.}
\begin{center}
\begin{tabular}{l|cc}
\toprule
Feature & 0s & 1s\\
\midrule
if\textunderscore fthb & 128,131 & 20,389\\
if\textunderscore prim\textunderscore res & 12,790 & 135,730\\
if\textunderscore corr & 89,602 & 58,918\\
if\textunderscore sf & 42,195 & 106,325\\
if\textunderscore purc & 95,380 & 53,140\\
if\textunderscore sc & 147,130 & 1,390\\
if\textunderscore delq\textunderscore sts & 118,184 & 30,336\\
default & 141,094 & 7,426\\
\bottomrule
\end{tabular}
\end{center}
\end{table}

\newpage

\section{Architecture of the DNN baseline model}
\label{Architecture of the DNN baseline model}

\begin{figure}[hbt!]
\includegraphics[scale=0.6]{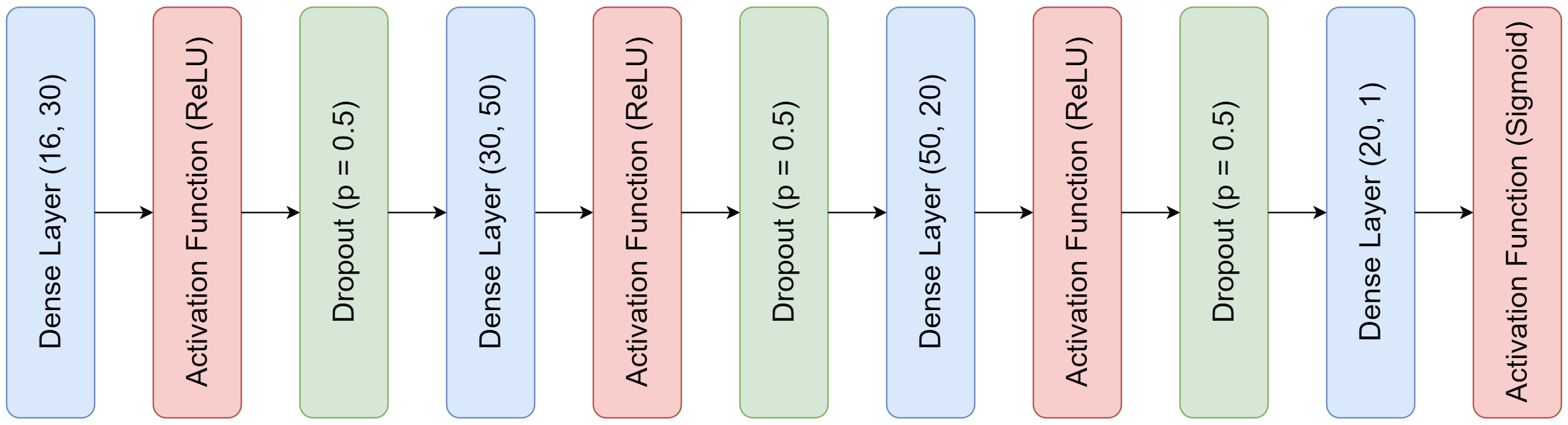}
\centering
\caption{Architecture of the DNN baseline model.}
\label{fig10}
\end{figure}

\section{Hyperparameters for model training and computation resources}
\label{Hyperparameters for model training and computation resources}

\begin{table}[hbt!]
\footnotesize
\caption{Hyperparameters for model training.}
\begin{center}
\begin{tabular}{l|l}
\toprule
Hyperparameter & Value\\
\midrule
Epochs & 200\\
Early stop & 50\\
Learning rate & 0.001\\
Optimizer & Adam\\
\bottomrule
\end{tabular}
\end{center}
\end{table}

\begin{table}[hbt!]
\footnotesize
\caption{Computation resources.}
\begin{center}
\begin{tabular}{l|l}
\toprule
Resource & Specification\\
\midrule
Processor & AMD Milan 7413 @ 2.65 GHz 128M cache L3\\
CPU cores per task & 2\\
GPU & NVidia A100\\
Memory per GPU & 40 GB\\
\bottomrule
\end{tabular}
\end{center}
\end{table}
\end{appendices}

\end{document}